\documentclass[
aps, pra,
twocolumn,
amsmath,amssymb,
shortbibliography,
superscriptaddress
]{revtex4-1}

\bibpunct{[}{]}{,}{n}{}{}

\usepackage[T1]{fontenc}
\usepackage[utf8]{inputenc}
\usepackage[english]{babel}

\usepackage{graphicx}
\usepackage{dcolumn}
\usepackage{bm}
\usepackage{braket}
\usepackage{color}
\usepackage[usenames,dvipsnames]{xcolor} 
\usepackage{newtxtext,newtxmath}
\usepackage[colorlinks=true,linkcolor=blue,citecolor=blue,urlcolor=blue]{hyperref} 

\usepackage{csquotes}
\usepackage{siunitx}
\usepackage{comment}

\newcommand{\jena}{Institut f\"ur Festk\"orpertheorie und -optik, Friedrich-Schiller-Universit\"at Jena, Max-Wien-Platz 1, 07743 Jena, Germany}
\newcommand{\etsf}{European Theoretical Spectroscopy Facility (ETSF)}
\newcommand{\coimbra}{CFisUC, Department of Physics, University of Coimbra, Rua Larga, 3004-516 Coimbra, Portugal}

\newcommand{\kv}{\mathbf{k}}
\newcommand{\kB}{k_\mathrm{B}}
\renewcommand{\Re}{\operatorname{Re}}
\renewcommand{\Im}{\operatorname{Im}}

\graphicspath{ {./Figs/} {./} }

\begin{document}

\title{
Ensemble averages of \emph{ab initio} optical, transport, and thermoelectric properties of hexagonal Si$_x$Ge$_{1-x}$ alloys}%
\author{Pedro Borlido}
\affiliation{\jena}
\affiliation{\etsf}
\affiliation{\coimbra}
\author{Friedhelm Bechstedt}
\author{Silvana Botti}
\author{Claudia R\"odl}
\affiliation{\jena}
\affiliation{\etsf}

\begin{abstract}
We present a comprehensive first-principles investigation of optical, transport, and thermoelectric properties of pure and doped hexagonal Si$_x$Ge$_{1-x}$ alloys based on density-functional theory calculations, the Boltzmann transport equation, and the generalized quasi-chemical approximation to obtain alloy averages of electronic properties. At low temperature, phase decomposition into the hexagonal elementary crystals is thermodynamically favored, but around and above room temperature random alloys are predicted to be stable. While hexagonal Si has an indirect band gap, the gap of hexagonal Ge is direct with very weak optical transitions at the absorption edge. The alloy band gap remains direct for a Si content below 45\,\% and the oscillator strength of the lowest optical transitions is efficiently enhanced by alloying. The optical spectra show clear trends and both absorption edges and prominent peaks can be tuned with composition. The dependence of transport coefficients on carrier concentration and temperature is similar in cubic and hexagonal alloys. However, the latter display anisotropic response due to the reduced hexagonal symmetry. In particular, the transport mass exhibits a significant directional dependence. Seebeck coefficients and thermoelectric power factors of $n$-doped alloys show non-monotonous variations with the Si content independently of temperature.
\end{abstract}
\maketitle


\section{Introduction}

Silicon (Si) and germanium (Ge) are outstandingly important materials for microelectronics. However, their indirect fundamental band gap in the cubic (cub) diamond crystal structure with space group $Fd\bar{3}m$  makes them inefficient light emitters and prevents their application to photonics. However, there is an increasing demand for on-chip integration of optoelectronic devices and achieving efficient light emission in Si-compatible materials has become an urgent need. Fortunately, modification of the stacking of Si planes in the (111) direction leads to another metastable phase: the hexagonal (hex) lonsdaleite structure with space group $P6_3/mmc$. The band structure of this Si phase yields a smaller but still indirect band gap~\cite{Raffy.Furthmueller.ea:2002:PRB,Roedl.Sander.ea:2015:PRB}, whereas the band gap becomes direct in the case of Ge~\cite{Roedl.Furthmueller.ea:2019:PRM,Borlido:PRM}. Consequently, hexagonal Si$_x$Ge$_{1-x}$ alloys have very promising tunable fundamental gaps and oscillator strengths~\cite{Fadaly:2020:N,Borlido:PRM}. Besides an increasing band gap, a direct-to-indirect transition is observable with rising Si molar fraction~$x$~\cite{Wang.Zhang.ea:2021:APL,Bao.Guo.ea:2021:JAP,Borlido:PRM}. Therefore, Ge-rich hex-Si$_x$Ge$_{1-x}$ alloys are emerging as a new class of optoelectronic materials with extraordinary properties for emerging photonic technologies.

By now, hex-Si$_x$Ge$_{1-x}$ with stacking fault volume fraction below 0.2\,\% can be synthesized~\cite{Fadaly:2020:N}. Hexagonal crystals of Si and Ge have been successfully produced using vapor-liquid-solid growth of nanowires or strain-induced phase transformation \cite{Fadaly:2020:N,Hauge.Verheijen.ea:2015:NL,Pandolfi.ReneroLecuna.ea:2018:NL,He.Maurice.ea:2019:N}. Moreover, hex-Si$_x$Ge$_{1-x}$ alloys have been grown epitaxially on wurtzite-GaAs or wurtzite-GaP nanowire cores~\cite{Fadaly:2020:N,Hauge.ConesaBoj.ea:2017:NL}. Tunable light emission has been achieved over a broad range of 0.3-0.7\,eV~\cite{Fadaly:2020:N}. Therefore, these hexagonal alloys have demonstrated to be a promising platform for Si-compatible photonic devices and intra-chip light communication technology.

The electronic properties of hexagonal Si$_x$Ge$_{1-x}$ alloys have been theoretically investigated by some authors based on a very restricted number of atomic configurations~\cite{Wang.Zhang.ea:2021:APL} using the special quasi-random structure (SQS) model~\cite{Bao.Guo.ea:2021:JAP} or the virtual-crystal approximation (VCA)~\cite{Cartoixa.Palummo.ea:2017:NL}. Note that the VCA band structures do not give a reliable description of the indirect-to-direct transition, as discussed in Ref.~\cite{Borlido:PRM}, where we presented the application of an extended statistics of atomic configurations for the determination of alloy-averaged electronic properties at high temperature. This statistical approach gives hexagonal lattice constants $P$, with $P=a$ or $P=c$, and $P(x)=x P_\mathrm{Si}+(1-x)P_\mathrm{Ge}-x(1-x)\Delta P$, that vary nearly linearly between $a=3.953$~{\AA} and $c=6.588$~{\AA} for hex-Ge and $a=3.826$~{\AA} and $c=6.327$~{\AA} for hex-Si, in good agreement with experimental data~\cite{Fadaly:2020:N,Hauge.ConesaBoj.ea:2017:NL}. The small bowing parameters $\Delta a=0.029$~{\AA} and $\Delta c=0.040$~{\AA} indicate minor deviations from Vegard's rule. Existing theoretical studies of optical properties are also based on a limited number of configurations~\cite{Bao.Guo.ea:2021:JAP,Cartoixa.Palummo.ea:2017:NL}. Evaluation of transport and thermoelectric properties of hexagonal alloys are still rare~\cite{Gu.Zhao:2018:JAP,hong2021n-type}, while more calculations are available for cubic SiGe alloys~\cite{xu2021first,hahn2021intrinsic,fujimoto2022thermoelectric}. 

Here, we fill the still existing gap of knowledge by presenting detailed \emph{ab initio} calculations for hex-Si$_x$Ge$_{1-x}$ alloys with varying Si content using state-of-the-art density-functional theory (DFT) and the generalized quasi-chemical approximation (GQCA) of alloy statistics. We determine the impact of alloying and chemical disorder on a variety of (anisotropic) optical spectra, transport coefficients, and thermoelectric functions. We also consider the dependence of transport and thermoelectric coefficients on free-carrier concentrations, thus accounting for effects of doping that are known to be present in experimental samples. In Section~\ref{sec:methods}, the underlying theoretical and computational approaches are summarized. Section~\ref{sec:energetics} briefly discusses the thermodynamic stability of hex-Si$_x$Ge$_{1-x}$ in different temperature regimes. Section~\ref{sec:optics} focuses on the optical spectra of the hexagonal alloys, whereas Section~\ref{sec:transport} discusses the transport and thermoelectric properties. A summary is given and conclusions are drawn in Section~\ref{sec:summary}.


\section{Theoretical and computational methods}
\label{sec:methods}

\subsection{Alloy statistics}

The properties of the hexagonal SiGe alloys were obtained by a thermodynamic ensemble average within the GQCA~\cite{Sher.Schilfgaarde.ea:1987:PRB,Teles.Furthmuller.ea:2000:PRB} that divides the alloy in independent clusters and takes the total energies of the individual clusters as well as the mixing entropy into account in the thermodynamic average. Within this approximation, a macroscopic binary alloy with composition Si$_x$Ge$_{1-x}$ and $N$ atomic sites is split into $M$ cluster cells with $n$ atomic sites each, such that $N=Mn$. For a $n$-atom cluster cell, there are $2^n$ possible arrangements of Si and Ge atoms. The cluster cells are grouped in $J+1$ different classes, with each class containing all clusters that are equivalent according to their space-group symmetry. By construction, every class $j$ ($j=0,\ldots,J$), with total energy per atom $E_j$, contains $g_j$ different atomic arrangements with $n_j$ Si atoms and $n-n_j$ Ge atoms ($\sum_{j=0}^J g_j=2^n$).

The set of non-equivalent clusters $\{M_0,M_1,\ldots,M_J\}$, with $\sum^J_{j=0}M_j=M$, describes how many clusters of each class occur in the alloy. A single class $j$ contributes to the macroscopic alloy with its cluster fraction $x_j=M_j/M$. The cluster fractions fulfill the constraints $\sum^J_{j=0}x_j=1$ and $\sum^J_{j=0}n_jx_j=nx$, with $x$ the molar fraction of Si atoms in the alloy. A composition-dependent property $P(x)$ of the Si$_x$Ge$_{1-x}$ alloy at composition $x$ is then obtained by a statistical average of the corresponding properties $P_j$ over all cluster classes according to the Connolly-Williams method \cite{Connolly.Williams:1983:PRB},
\begin{equation}
   \langle P(x) \rangle = \sum^J_{j=0} x_j P_j.
   \label{eq:connoly-average}
\end{equation}

In general, the cluster fractions 
\begin{equation}
    x_j = \frac{ g_j \eta^{n_j} \exp\left( -\beta \Delta E_j \right) }{ \sum_k g_k \eta^{n_k} \exp\left( - \beta \Delta E_k \right) },
    \label{eq:cell_fractions}
\end{equation}
(with $\beta = 1/(k_\text{B} T)$), depend on both temperature $T$ and the excess energies per atom
\begin{equation}
   \Delta E_j = \frac{E_j}{n}  - \frac{n_j}{n} E_\text{Si} - \frac{n-n_j}{n}  E_\text{Ge},
   \label{eq:excess-energies}
\end{equation}
where $E_\text{Si}$ and $E_\text{Ge}$ are the total energies per atom of lonsdaleite Si and Ge, and the coefficient $\eta$ is found by minimization of the Helmholtz free energy with the above constraints. 

In the high-temperature limit $T \rightarrow \infty$, the cluster fractions $x_j$ become
\begin{equation}
   x_j^0 = g_j\, x^{n_j}(1-x)^{n-n_j}
   \label{eq:srs-fractions}
\end{equation}
and the dependency on the excess energies disappears. In turn, the configuration entropy per atom reduces to
\begin{equation}
    \Delta S = -\kB \left[ x\ln(x) + (1-x) \ln(1-x)  \right],
    \label{eq:srs-entropy}
\end{equation}
resulting in a purely stochastic distribution. The high-temperature limit is known as strict-regular solution (SRS) model~ \cite{Sher.Schilfgaarde.ea:1987:PRB,Schleife.Eisenacher.ea:2010:PRB}. 

We use the \textsc{genstr} tool of the Alloy Theoretical Automated Toolkit (\textsc{atat}) \cite{vandeWalle.Asta.ea:2002:C} to generate all possible $8$-atom clusters resulting in 118 symmetry-inequivalent cluster classes of three different cell shapes. This involves clusters that account for chemical disorder both perpendicular and along the hexagonal axis. Using \textsc{atat}, we can also obtain alloy total energies from a cluster expansion that relies on fitting a Heisenberg model to the total energies of the cluster cells. This represents an alternative approach to evaluate the phase diagram of the alloy and yields important information regarding the convergence of the calculations with the cluster-cell size.


\subsection{Total energy and electronic structure calculations}

The ground-state properties of the individual SiGe clusters were calculated in the framework of DFT as implementend in \textsc{VASP}~\cite{Kresse.Furthmueller:1996:PRB,Kresse.Furthmueller:1996:CMS} using the projector-augmented wave method~\cite{Kresse.Joubert:1999:PRB} with a plane-wave cutoff of 500~eV. The shallow Ge\,$3d$ electrons are treated as valence states.

For geometry optimization and the computation of total energies, we employ the PBEsol exchange-correlation functional~\cite{Perdew.Burke.ea:1996,Perdew.Ruzsinszky.ea:2008:PRL}, that has been shown to yield accurate lattice parameters for bulk solids~\cite{Csonka.Perdew.ea:2009:PRB,Zhang.Reilly.ea:2018:NJoP}, including the cubic and hexagonal phases of Si and Ge~\cite{Roedl.Furthmueller.ea:2019:PRM,Borlido:PRM,Suckert.Roedl.ea:2021:PRM}. Brillouin-zone integrations were performed using $\Gamma$-centered $\kv$-point grids, with a $\kv$-point density equivalent to a $12\times12\times6$ mesh for the primitive londsdaleite cell. The atomic geometries were relaxed until the forces acting on the atoms drop below 1~meV/{\AA}.

In order to accurately describe the electronic structure of the individual clusters, we used the MBJLDA meta-GGA functional proposed by Tran and Blaha \cite{Tran.Blaha.ea:2007:JoPCM,Tran.Blaha:2009:PRL} that is based on a modified Becke-Johnson potential \cite{Becke.Johnson:2006:TJoCP}. Spin-orbit coupling is always included, as the resulting corrections to the band structure are important for alloys with large Ge content. The MBJLDA functional has been shown to yield excellent agreement with experimental band gaps and band structures for cub-Si and cub-Ge \cite{Laubscher.Kuefner.ea:2015:JoPCM}, hex-Si~\cite{Borlido:PRM}, hex-Ge~\cite{Roedl.Furthmueller.ea:2019:PRM,Borlido:PRM}, and generally for semiconductors~\cite{Kim.Hummer.ea:2009:PRB,Borlido.Aull.ea:2019:JCTC}. 


\subsection{Optical spectra}

The frequency-dependent dielectric function is calculated based on the MBJLDA electronic structure within the independent-particle approximation (see e.g.\ Ref.~\cite{Bechstedt:2015:Book}),
\begin{multline}
\varepsilon_{ii}(\omega)=1-\frac{e^2\hbar^2}{\varepsilon_0 m_\text{e}^2}\frac{1}{\Omega_0}\sum_{\nu\nu'\kv} w_\kv \\
\times\frac{|\braket{\nu\kv|p_i|\nu'\kv}|^2}{\left(\varepsilon_{\nu\kv}-\varepsilon_{\nu'\kv}\right)^2} 
\frac{f(\varepsilon_{\nu'\kv})-f(\varepsilon_{\nu\kv})}{\varepsilon_{\nu'\kv}-\varepsilon_{\nu\kv}+\hbar(\omega+i\gamma)},
	\label{eq:ipa-dielectric-tensor}
\end{multline}
are the Fermi occupation numbers at temperature $T=0$, and $p_i$ the Cartesian components of the momentum operator. The one-particle eigenstates $\ket{\nu\kv}$ with eigenvalues $\varepsilon_{\nu\kv}$ are labeled by the band index $\nu$ and the $\kv$~point. The $\kv$-point weight for Brillouin-zone integrations is given by $w_\kv$ and $\Omega_0$ is the unit-cell volume. Excitonic effects are neglected, as the exciton binding energies of group-IV materials are very small (of the order of a a few meV \cite{MACFARLANE1959388,Altarelli.Lipari:1976:PRL}) due to strong dielectric screening. We remark that excitonic effects, however, are expected to modify the spectral shape. Optical properties for the full set of 8-atom cluster cells were calculated on $\kv$-point grids with a density equivalent to $24\times24\times12$ points for the primitive lonsdaleite cell and a Lorentzian broadening of $\gamma=0.1$~eV. With this setting, we can sample the effect of chemical disorder over the entire frequency range of optical excitations.

For a detailed analysis of the absorption edge, a considerably lower broadening, and consequently much denser $\kv$-point grids, are required. Due to the large number of alloy configurations and low symmetries of the 8-atom cluster cells, this is prohibitive for the entire set of clusters. Instead, we additionally computed the optical properties in the vicinity of the absorption edge for the much smaller set of 4-atom cluster cells on a denser grid equivalent to $36\times36\times18$ $\kv$~points for the primitive lonsdaleite cell. 

As the inverse lifetime at the absorption edge is orders of magnitude lower than the Lorentzian broadening of $\gamma=0.1$~eV, this leads to artificially strong Lorentzian tails within the band gap which become very prominent in logarithmic plots of the absorption coeffcient. This problem can be overcome by replacing the Lorentzian representation of the $\delta$ functions in $\Im\varepsilon(\omega)$ by the bump function \cite{Roedl.Sander.ea:2015:PRB}, a nascent $\delta$~function with compact support,
\begin{equation} \label{eq:bump}
B_b(\omega) = 
\begin{cases}
\frac{1}{cb} \exp\left( \frac{b^2}{\omega^2 - b^2} \right) & \text{for }|\omega|<b \\
0 & \text{otherwise}
\end{cases}.
\end{equation}
Here, $c=0.44399382$ is a normalization constant and the broadening parameter $b$ is adjusted such that the height of the bump corresponds to the height of a Lorentzian with $\gamma=0.05$~eV. See Supp.\ Mat.\ for a detailed comparison between bump and Lorentzian broadening.

The alloy-averaged dielectric function $\left< \varepsilon_{ii} (\omega) \right>$ is obtained by a pointwise alloy average of $\varepsilon_{ii}(\omega)$ at each frequency $\omega$ according to Eq.~\eqref{eq:srs-fractions}. Defining the alloy average of optical properties in this way is justified, since optical measurements probe the average atomic structure of the material over the length scale of optical wavelengths, which is significantly larger than the length scale of structural disorder in the alloy. The refractive index $\langle n_{ii}(\omega)\rangle$
and extinction coefficient $\langle\kappa_{ii}(\omega)\rangle$ of the alloy are evaluated according to
\begin{equation}
    \left < n_{ii}(\omega) \right >
    =
    \sqrt{ 
        \frac{ | \langle \varepsilon_{ii}(\omega) \rangle | + \Re \langle \varepsilon_{ii}(\omega) \rangle }{2}
    }
	\label{eq:refractive-index-alloy}
\end{equation}
and
\begin{equation}
    \left < \kappa_{ii}(\omega) \right >
    =
    \sqrt{ 
        \frac{ | \langle \varepsilon_{ii}(\omega) \rangle | - \Re \langle \varepsilon_{ii}(\omega) \rangle }{2}
    }.
	\label{eq:extinction-coeff-alloy}
\end{equation}
Consequently, the reflectivity at normal incidence reads
\begin{equation}
	\langle R_{ii} (\omega)\rangle = \frac{\left[1-\langle n_{ii} (\omega)\rangle\right]^2+
	\langle\kappa_{ii} (\omega)\rangle^2}{\left[1+\langle n_{ii} (\omega)\rangle\right]^2+
	\langle\kappa_{ii} (\omega)\rangle^2}
	\label{eq:reflectivity}
\end{equation}
and the absorption coefficient
\begin{equation}
	\langle\alpha_{ii} (\omega)\rangle  = \frac{2\omega}{c}\langle\kappa_{ii} (\omega)\rangle.
	\label{eq:absorption-coefficient}
\end{equation}


\subsection{Transport and thermoelectric properties}

In view of the interest of hex-Si$_x$Ge$_{1-x}$ for optoelectronic applications, we calculated transport properties using the linearized Boltzmann transport equation~\cite{Ashcroft.Mermin:1976:Book}, within the approximations of rigid bands and constant relaxation times, to deduce transport and thermoelectric coefficients. By means of the \textsc{Boltztrap2} package~\cite{MADSEN2018140}, the one-particle eigenvalues used for the calculation of the optical spectra were interpolated to grids containing five times more $\kv$ points in the Brillouin zone, in order to evaluate the conductivity tensor $\sigma_{ij}$, the Seebeck coefficient $S$, and the electronic contribution to the thermal conductivity $\kappa_\textrm{e}$. Additionally, we computed the power factor $S\sigma^2$. Based on the Drude model of conductivity $\sigma_{ij}$, we can also define an effective transport mass tensor~\cite{Ashcroft.Mermin:1976:Book,hautierChem.Mater.2014},
\begin{equation}
   \overline{m}^{-1}_{ij} = \frac{ \sigma_{ij} }{ \rho e^2 \tau } \, ,
   \label{eq:transport-effective-mass-tensor}
\end{equation}
where $\tau$ is the relaxation time, assumed to be isotropic and wave-vector independent, and $\rho$ is the charge-carrier concentration. The diagonal components of  $\overline{m}^{-1}_{ij}$ are the inverse transport effective masses averaged over the Brillouin-zone regions and bands contributing to the charge current, as a function of the position of the chemical potential.

Within these approximations, the relaxation time is an empirical parameter that can be further approximated using material-specific empirical models for different scattering processes. 
For an electron gas with electron densities of $3\ldots4\cdot10^{22}$\,cm$^{-3}$, $\tau$ is typically of the order of $10^{-14}$ to $10^{-15}$~s~\cite{Ashcroft.Mermin:1976:Book}. At much lower carrier densities, in doped semiconductors, an increase of $\tau$ by orders of magnitude can be expected. Unfortunately, an accurate evaluation of $\tau$ from first principles is extremely hard and expensive to perform, as it requires calculations involving electron-phonon coupling, interaction with defects, impurities, alloy disorder, etc. A more efficient alternative often favored in the literature is to compute $\tau$ using the deformation-potential method of Bardeen and Shockley~\cite{PhysRev.80.72}, which considers deformation potentials for intra-valley scattering by longitudinal acoustic phonons. This approach gives a rough estimate of the relaxation time, but does not include the contribution of alloy scattering. Having in mind the limitations of this approximation, we performed calculations of the input parameters of the model, using cluster cells with 4 atoms in the unit cell (see Supp.\ Mat.\ for more details). Overall, we conclude that the relaxation time is of the order of $500$~\si{\femto\second} (at $300$~\si{\kelvin}) throughout the whole composition range. This value is close to the one of cubic SiGe alloys, making us believe that the estimate is reliable. However, we prefer to keep the discussion independent of the value of $\tau$ and therefore present the results in terms of ratios of the transport coefficients and $\tau$ whenever necessary.


\section{Energetics}
\label{sec:energetics}

\subsection{Alloy average}

\begin{figure}
	\includegraphics[width=\linewidth]{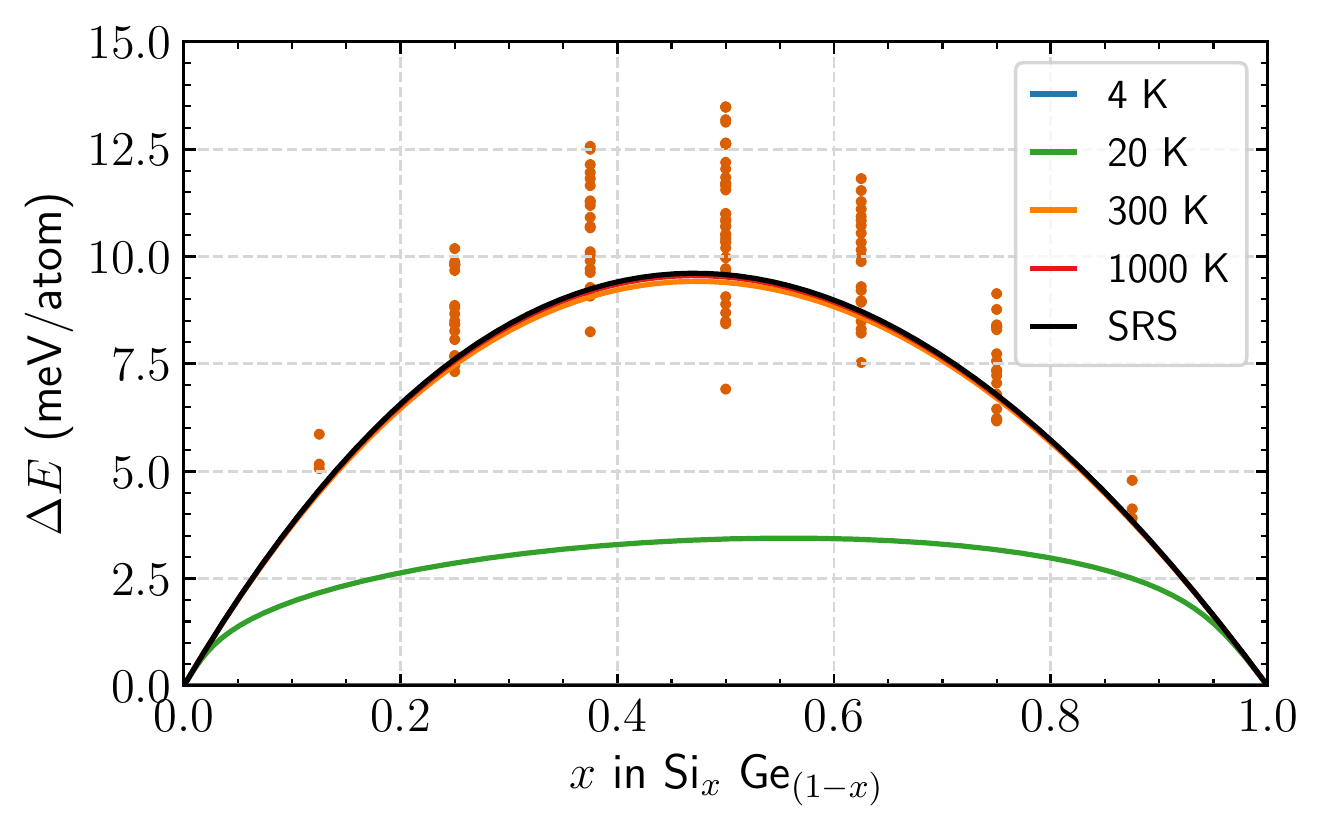}
	\caption{Excess energy per atom of hex-Si$_{x}$Ge$_{1-x}$ as a function of composition. Orange dots represent the excess energies of all clusters with 8 atoms in the unit cell. The lines show the alloy-averaged excess energies $\langle \Delta E \rangle$  at different temperatures in the GQCA and the SRS high-temperature limit. }
	\label{fig:excess-energy-sige}
\end{figure}

We first evaluate the accuracy of the different approximations to the thermodynamic alloy average. To this end, the excess energies per atom $\Delta E_j$ of all 8-atom clusters $j$ are shown in Fig.~\ref{fig:excess-energy-sige}, along with the thermodynamic average in the GQCA for various temperatures and the SRS high-temperature limit. It is obvious that the SRS limit is a very good approximation to the more sophisticated GQCA at room temperature and even more so at the growth temperature of hex-Si$_x$Ge$_{1-x}$ of about 1000~K~\cite{Fadaly:2020:N}. Only at very low temperatures, taking the energies of the clusters into account when evaluating the cluster fractions according to Eq.~\eqref{eq:cell_fractions} plays a non-negligible role and yields curves that deviate strongly from the SRS limit, until they become an almost linear interpolation between the end components of the alloy for $T\to0$. 

The maximum excess energy per atom of the 8-atom cells is about 14~meV/atom~ (see Fig.~\ref{fig:excess-energy-sige} and Ref.~\cite{Borlido:PRM}), considerably smaller than the thermal energy $\kB T$ at the typical growth temperatures. This explains why the cluster energies do not have a significant impact on the cluster fractions and the alloy can be considered random. We conclude that approximating the thermodynamic alloy average with the SRS limit is well justified. This holds also when optical measurements~\cite{Fadaly:2020:N} are performed at low temperature, as the structure does not change. Therefore, we will calculate all averages (except those required for the phase diagrams) in the SRS high-temperature limit which corresponds to the experimental growth conditions.


\subsection{Formation energies}

\begin{figure}
	\includegraphics[width=\linewidth]{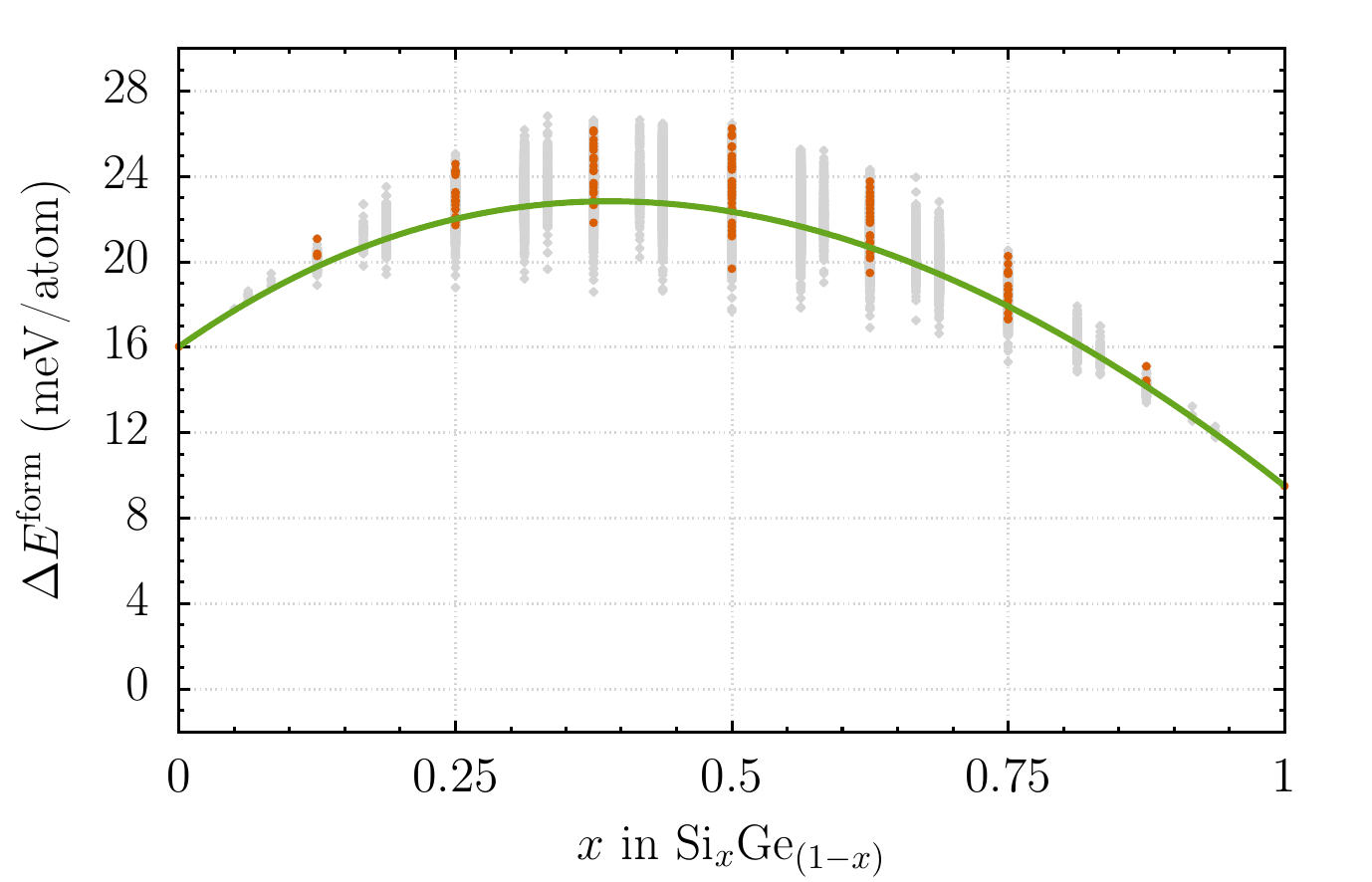}
	\caption{Formation energy per atom of hex-Si$_{x}$Ge$_{1-x}$ as a function of composition. The energy zeros are set to the formation energies of cub-Si and cub-Ge. Orange dots (gray diamonds) represent the values calculated with DFT (ATAT) for the set of inequivalent clusters with up to 8 (12) atoms in the unit cell. The green line represents the SRS average for the set of 8-atom clusters.} 
	\label{fig:formation-energy-sige}
\end{figure}

In Fig.~\ref{fig:formation-energy-sige}, we plot the formation energy $\Delta E_j^\mathrm{form}$  per atom for each cluster cell as a function of its stoichiometry $n_j/n$. The formation energy is defined in analogy to the excess energies in Eq.~\eqref{eq:excess-energies}, but with respect to the total energies per atom of Si and Ge in the diamond structure, the lowest-energy polymorphs of elementary Si and Ge. The alloy-averaged formation energy per atom within the SRS model is also presented for all alloy compositions.

Similarly to the alloy-averaged excess energy per atom (see Fig.~\ref{fig:excess-energy-sige} or Ref.~\onlinecite{Borlido:PRM}), the formation energy per atom is asymmetric with respect to the composition $x$. However, the asymmetry of the excess energy is much less pronounced than the one of the formation energy. The maximum of the alloy-averaged excess energy of $\left< \Delta E \right>(x)=10$~meV/atom is located at $x=0.47$~\cite{Borlido:PRM}. This small positive value is compatible with values of $\Delta E(x=0.5)=9$~meV/atom for cubic Si$_x$Ge$_{1-x}$ alloys~\cite{Martins.Zunger:1986:PRL}.
The maximum of the formation energy $\Delta E^\mathrm{form}=23$~meV/atom is located at $x=0.39$. The difference between these energies is explained by the different formation energies of the elementary solids in the hexagonal crystal structure. We find a formation energy of $10$~meV/atom for hex-Si and $16$~meV/atom for hex-Ge, in agreement with other \emph{ab initio} computations~\cite{Raffy.Furthmueller.ea:2002:PRB}.

The concavity of the formation-energy curve over alloy composition reflects the fact that the mixing of Si and Ge atoms is energetically unfavorable, which is known from cubic SiGe alloys. Because of the isovalency of the atoms, this can be traced back to the large difference of Si-Si and Ge-Ge bond lengths (about 4\,\%), which induces internal strain in the mixed crystals \cite{Martins.Zunger:1987:InProc}. 


\subsection{Phase diagram}

\begin{figure}
    \includegraphics[width=\linewidth]{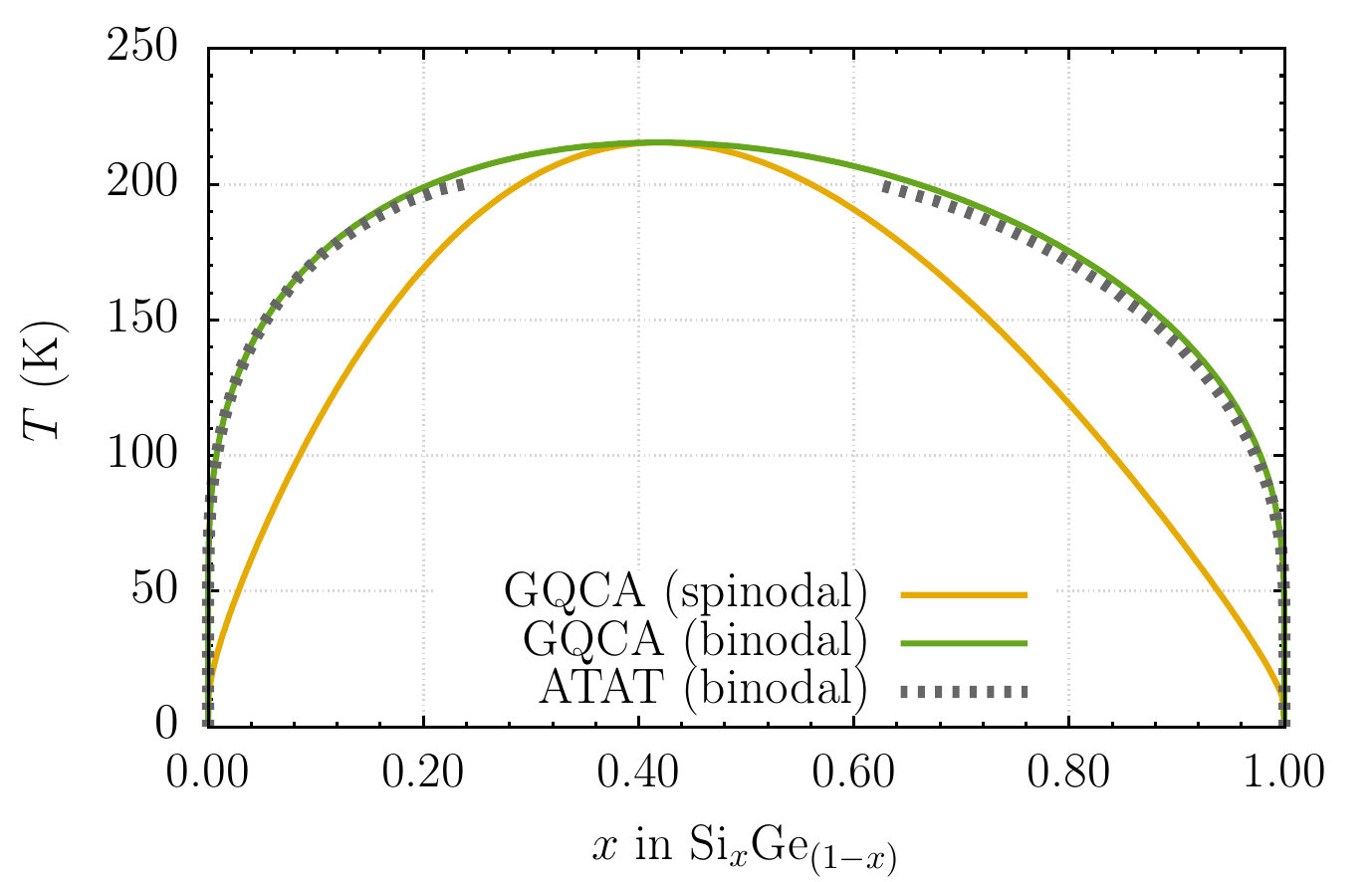} 
    \caption{Temperature-composition phase diagram of hexagonal Si$_x$Ge$_{1-x}$ alloys. The binodal line (green solid line) and the spinodal line (yellow solid line) are derived within the GQCA. For comparison, the binodal phase boundary as derived from Monte-Carlo calculations with \textsc{atat} (dotted line) is also given. }
    \label{fig:sige-t-vs-x-phase-diagram}
\end{figure}

The absence of ordered alloys with negative excess energies implies that hexagonal SiGe alloys are unstable with respect to decomposition in the elementary phases at zero temperature, and we expect to see a miscibility gap in the temperature-composition phase diagram. To elucidate this, it is useful to consider the evolution of the mixing free energy $\Delta F(x,T)=\Delta E(x)-T\Delta S(x)$ as a function of temperature. If $\Delta E>0$, then, as the temperature increases and the entropic term starts to dominate, a range of compositions exists where the decomposition into a silicon-rich and a germanium-rich phase lowers the free energy of the system. Using the common tangent-line method, we find the binodal lines given by the concentrations $x_1(T)$ and $x_2(T)$ that define the boundary between the separated phases and a fully mixed system in the $T$-$x$ phase diagram. Note that $\Delta E >0$ does not preclude the existence of ordered alloys. For example, in the case of strained cubic SiGe alloys, an order-disorder transition leading to a superlattice along the [111] axis has been experimentally observed~\cite{Ourmazd.Bean:1985:PRL}. The spinodal lines $x'_1(T)$ and $x'_2(T)$ are given by the inflection points of the free energy $\Delta F(x,T)$ as a function of composition at fixed temperature $T$. For compositions in the intervals $x_1<x<x'_1$ and $x'_2<x<x_2$, the alloy is metastable against local decomposition thanks to energy barriers.

Since the GQCA provides an analytical expression for $\Delta F$, it is easy to extract the $T$-$x$ phase diagram. Alternatively, one can calculate $\Delta F$ from Monte-Carlo simulations using a generalized Heisenberg model, as implemented in \textsc{atat}. We used both of these approaches to draw the phase diagram of hex-Si$_x$Ge$_{1-x}$ shown in Fig.~\ref{fig:sige-t-vs-x-phase-diagram}. The phase diagram obtained with \textsc{atat} is not complete and exhibits a gap close to the maximum of the curve. In this region of the diagram, Monte-Carlo simulations critically slow down, due to the difficulty of finding new configurations that are statistically independent of the previous ones. Although it is possible to bypass this problem, the overall good agreement with the GQCA curve and the additional computational cost made this unappealing. Instead, we interpolated the Monte-Carlo results within the gap via a 4-th order polynomial.

The GQCA predicts a critical temperature $T_\text{crit} = 215$~K at a critical composition $x_\text{crit} = 0.42$. From the \textsc{atat} curve, we extrapolated a similar critical point, with  $T_\text{crit} = 212$~K and $x_\text{crit} = 0.42$. This critical temperature is compatible with the observation that Si and Ge always mix at the experimental growth temperatures well above 300~K. The same holds for cubic SiGe alloys~\cite{Schaeffler:2001:InBook}.


\section{Optical properties}
\label{sec:optics}

\subsection{Dielectric function}

\begin{figure*}
    \centering
    \includegraphics[width=0.48\linewidth]{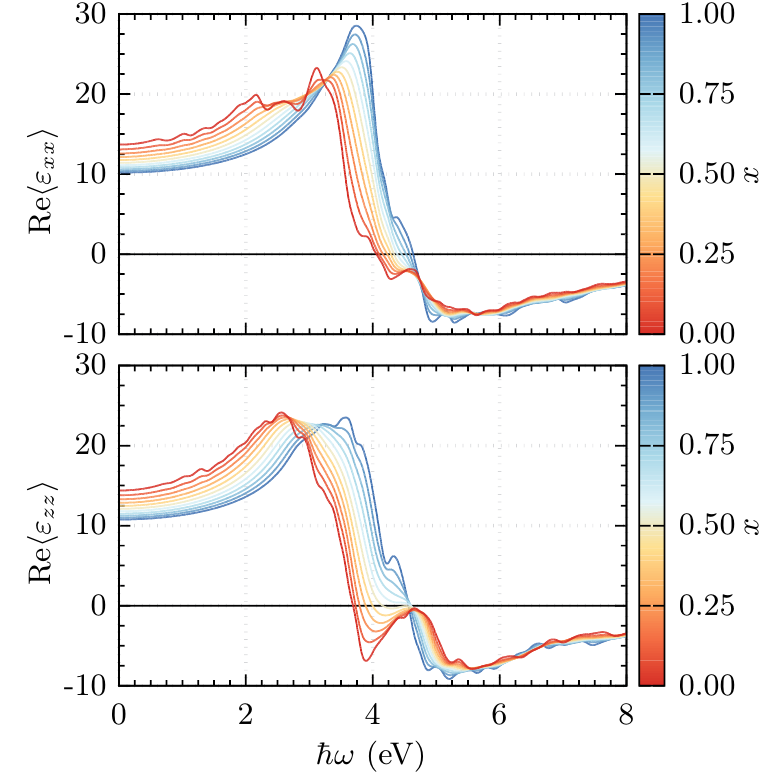} 
    \includegraphics[width=0.48\linewidth]{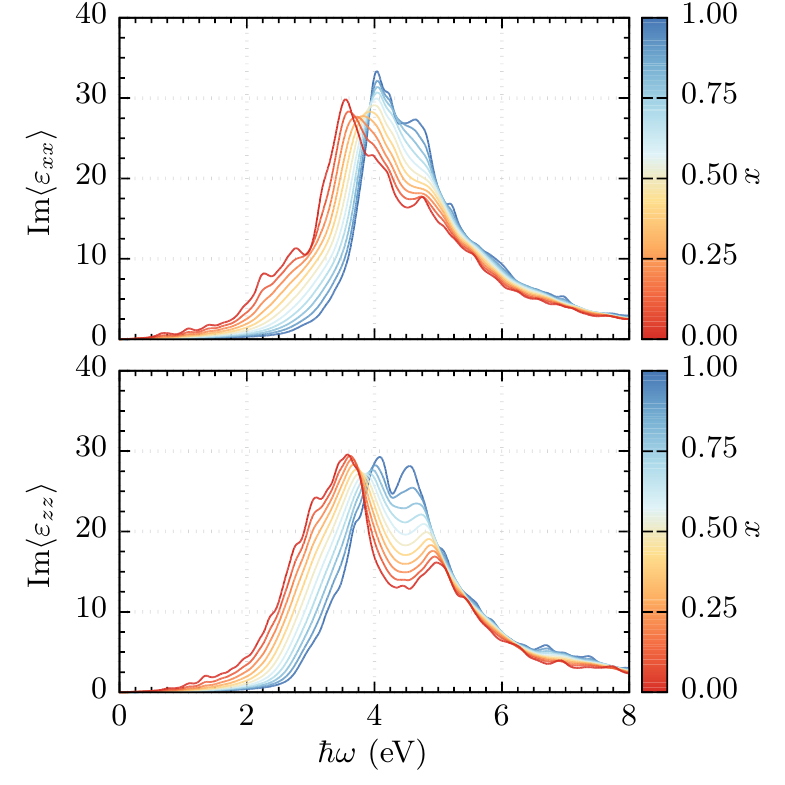}
   \caption{Evolution of the alloy-averaged real (left) and imaginary (right) parts of the dielectric function $\langle \varepsilon_{ii}(\omega) \rangle$ with alloy composition $x$. The average is obtained from the full set of 8-atom cluster cells. Results for in-plane (top) and out-of-plane (bottom) light polarization are shown. }
    \label{fig:alloy-dielectric-function}
\end{figure*}

The evolution of the real and imaginary parts of the alloy-averaged dielectric function as a function of composition is presented in Fig.~\ref{fig:alloy-dielectric-function}. We observe that the electronic static dielectric constant decreases as the Si concentration $x$ increases. For the in-plane (out-of-plane) components, the values range from $13.7$ ($14.4$) for hex-Ge to $10.2$ ($10.8$) for hex-Si, in agreement with previously reported results obtained with special quasi-random supercells~\cite{Bao.Guo.ea:2021:JAP}. Along with the increasing band gap, the dominant absorption peaks move towards higher energies for increasing Si content.

In general, alloying does not induce any additional features in the spectra, as peak positions and relative intensities vary smoothly with composition. In analogy with the cubic alloy, the two main peaks in the spectra can be labelled $E_1$ and $E_2$ and associated to the van-Hove singularities observed in cub-Si, cub-Ge~\cite{Yu.Cardona:2010:Book}, and cubic Si$_x$Ge$_{1-x}$ alloys \cite{Humlicek.Garriga.ea:1989:JoAP}. In hexagonal materials, these peaks are less pronounced and additionally depend on the polarization direction. From band-folding arguments, the peak evolving with composition from 3.6~eV (hex-Ge) to 4.0~eV (hex-Si) can be related to the $E_1$ van-Hove singularity of cubic Si and Ge~\cite{Humlicek.Garriga.ea:1989:JoAP}. However, the strong blueshift of the low-energy peak position when going from pure Ge to pure Si does not occur in the hexagonal case. The higher-energy peak near 5~eV in hex-Ge shifts to slightly lower photon energies of about 4.6~eV in hex-Si. The behavior of this peak, including the higher peak intensity for hex-Si, is similar to what is observed in cubic SiGe alloys.

We conclude that hexagonal SiGe alloys behave qualitatively similar to their cubic counterparts (see e.g.\ Refs.~\cite{ahujaJournalofAppliedPhysics2003,bahngJ.Phys.:Condens.Matter2001,Humlicek.Garriga.ea:1989:JoAP,jellisonOpticalMaterials1993}). In addition, the roots of $\Re \langle \varepsilon \rangle$, which lie within the range of $3.7$ to $4.6$~\si{\eV}, are close to those found for cubic Si$_x$Ge$_{1-x}$ \cite{Humlicek.Garriga.ea:1989:JoAP}. They may be interpreted in terms of Penn gaps of the averaged material \cite{Bechstedt:2015:Book}, slightly affected by the optical anisotropy.


\subsection{Absorption edge}

\begin{figure*}
	\begin{tabular}{c c}
		   \includegraphics[width=0.45\linewidth]{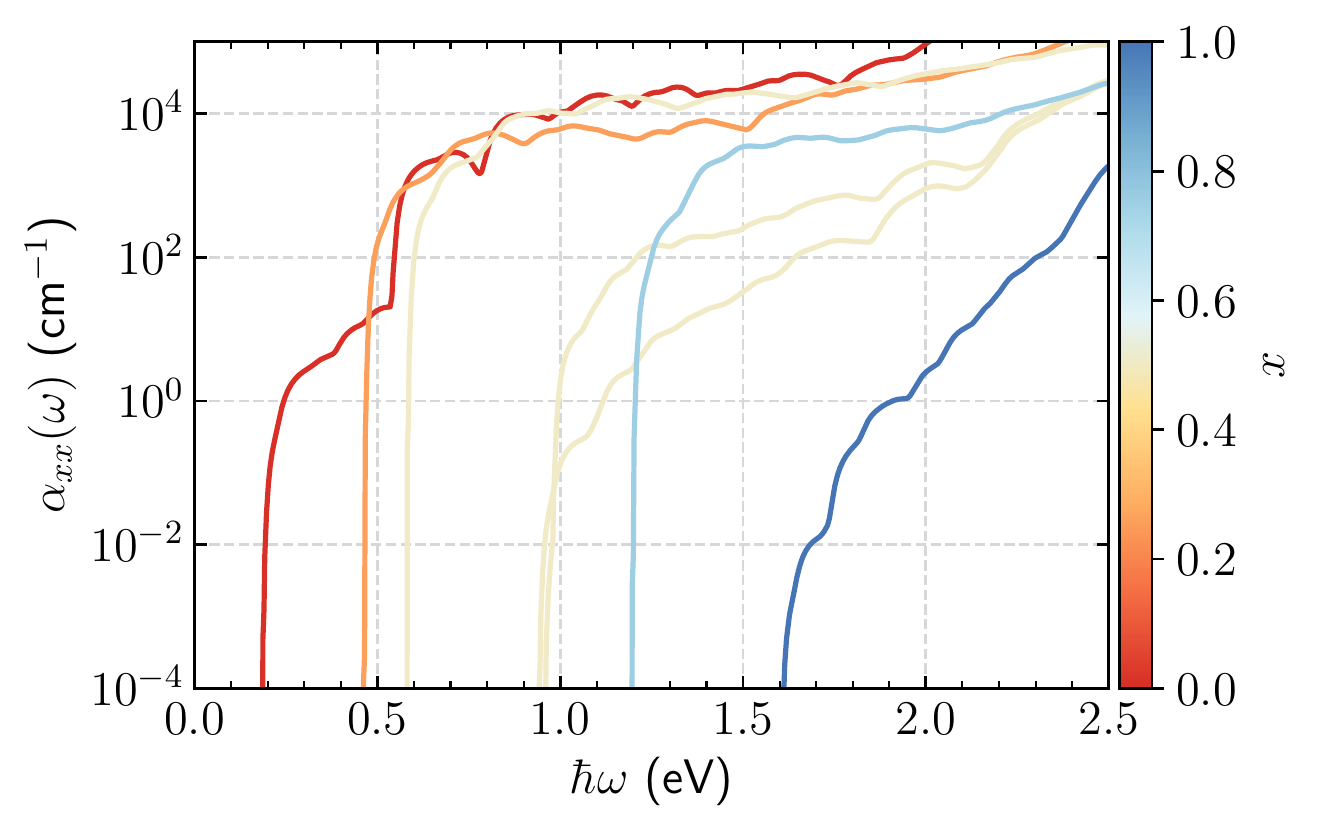} &
		   \includegraphics[width=0.45\linewidth]{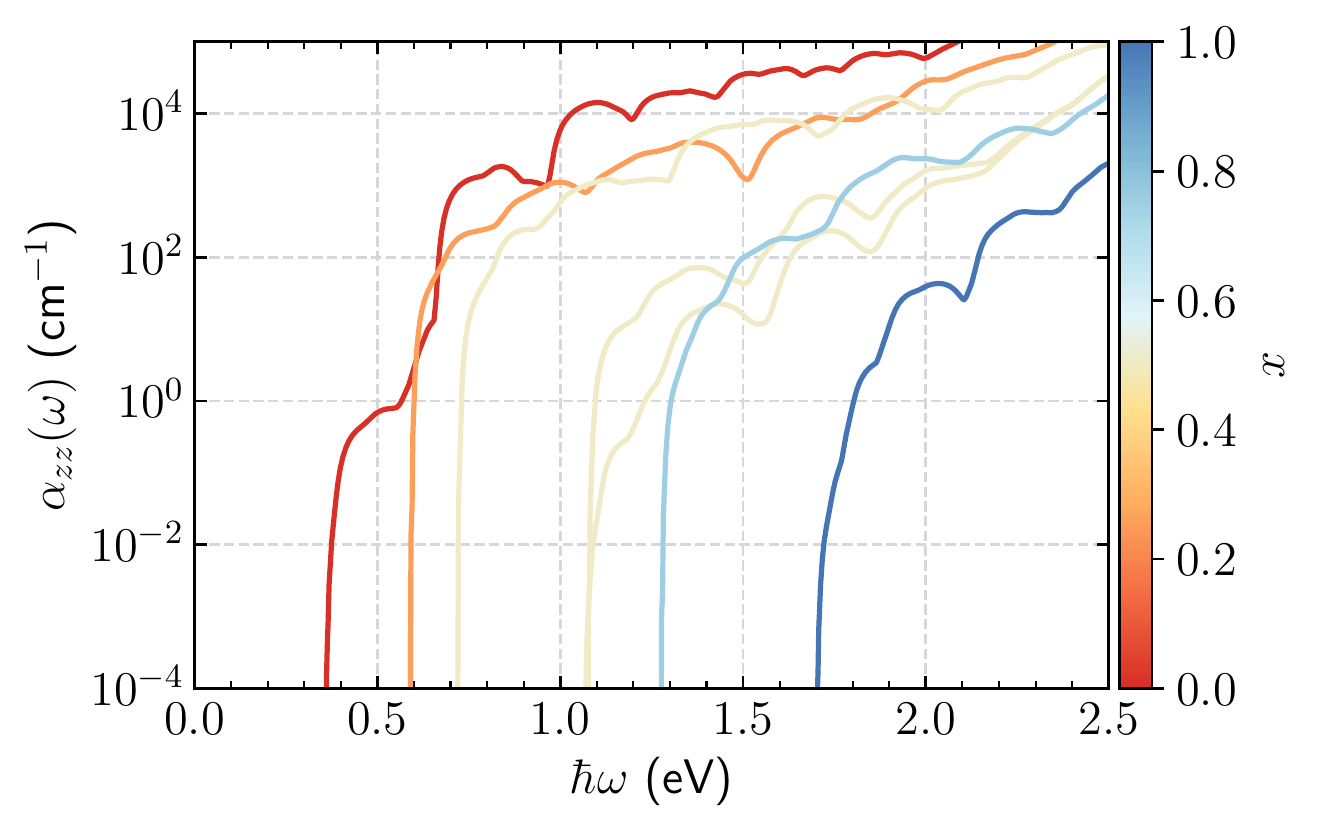} \\
		   \includegraphics[width=0.45\linewidth]{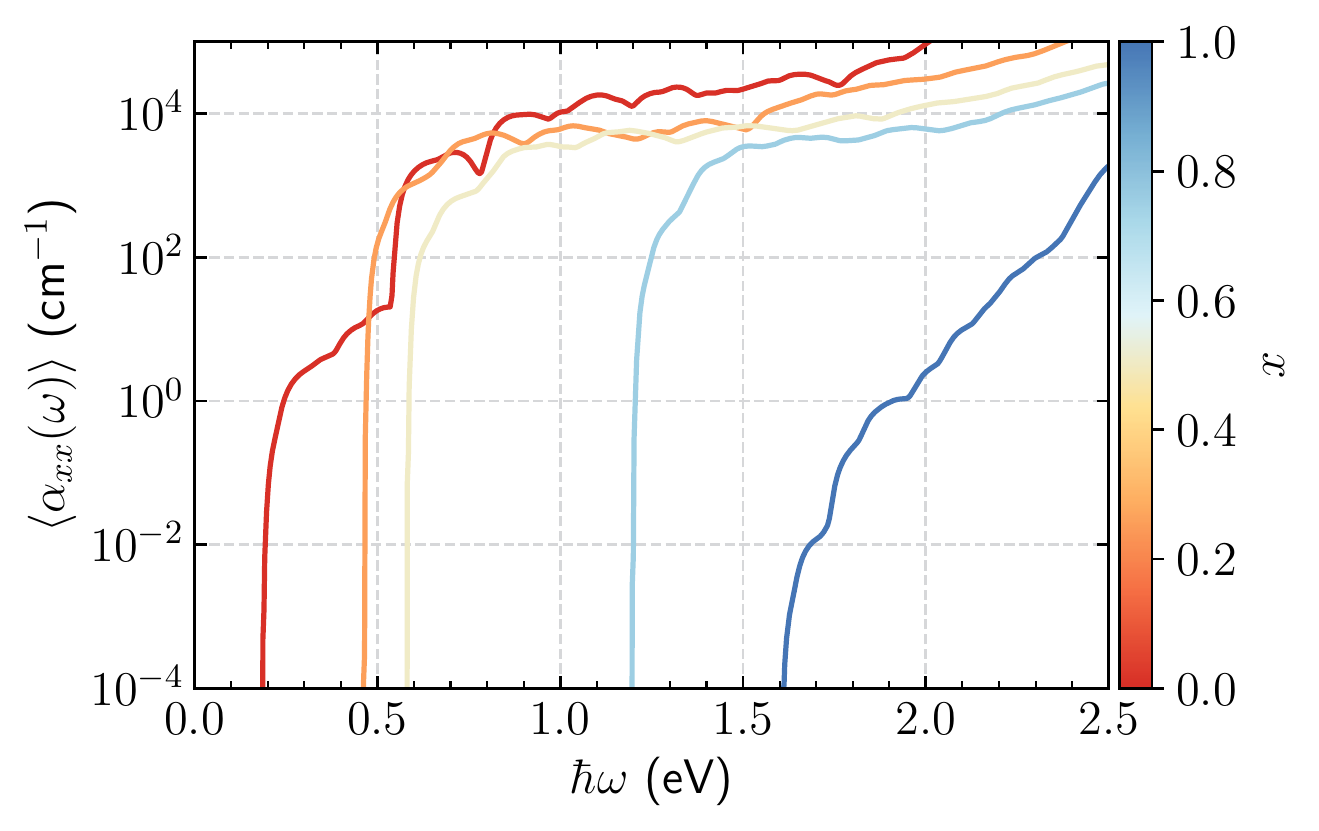} &
		   \includegraphics[width=0.45\linewidth]{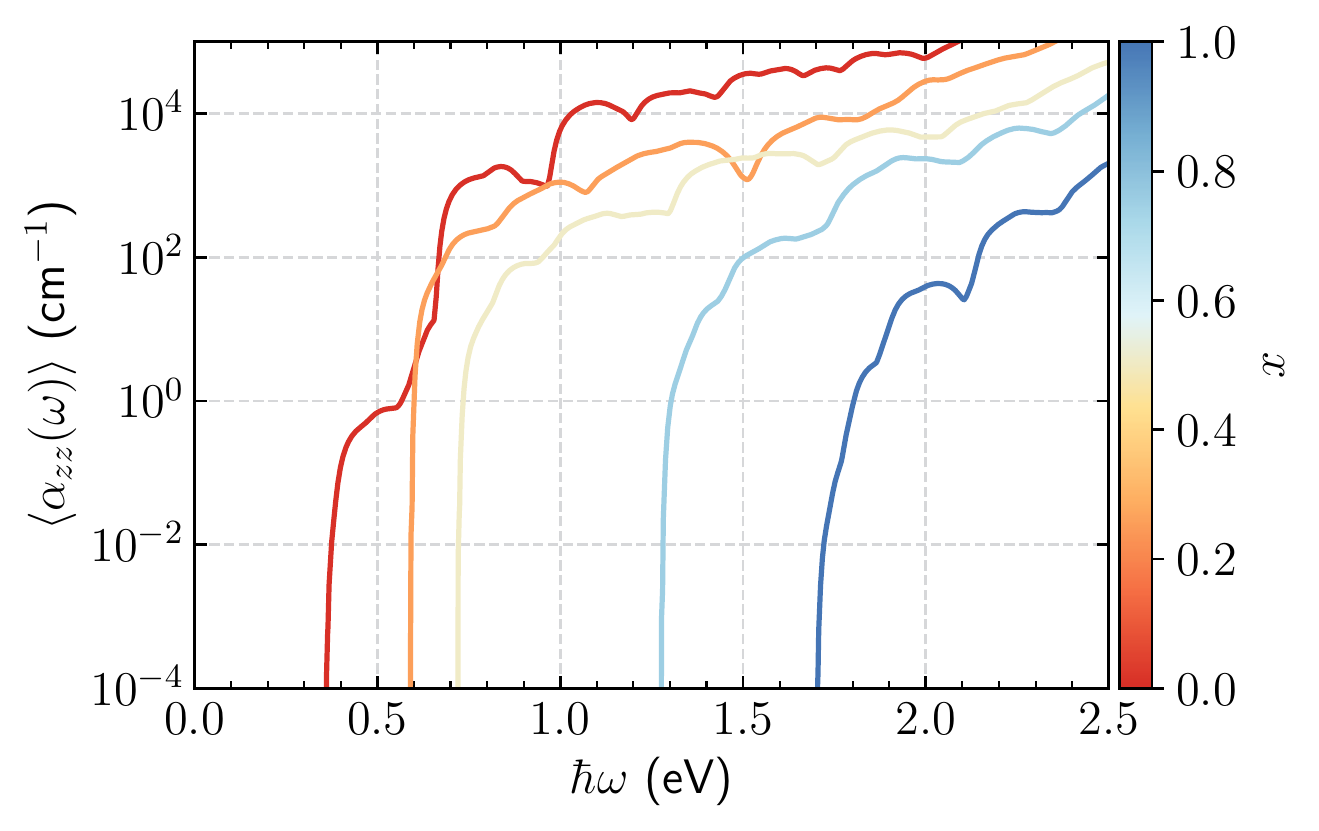} \\
		   \includegraphics[width=0.45\linewidth]{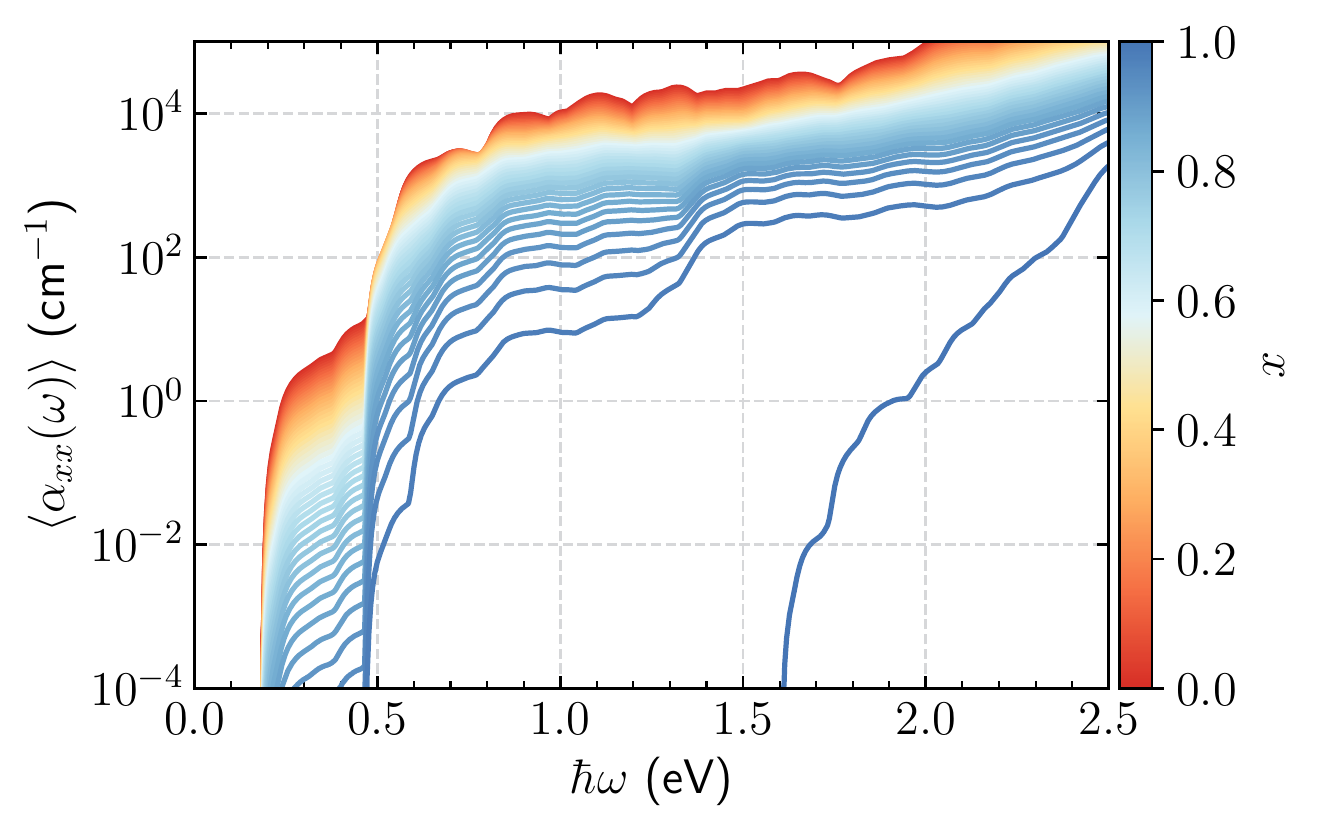} &
		   \includegraphics[width=0.45\linewidth]{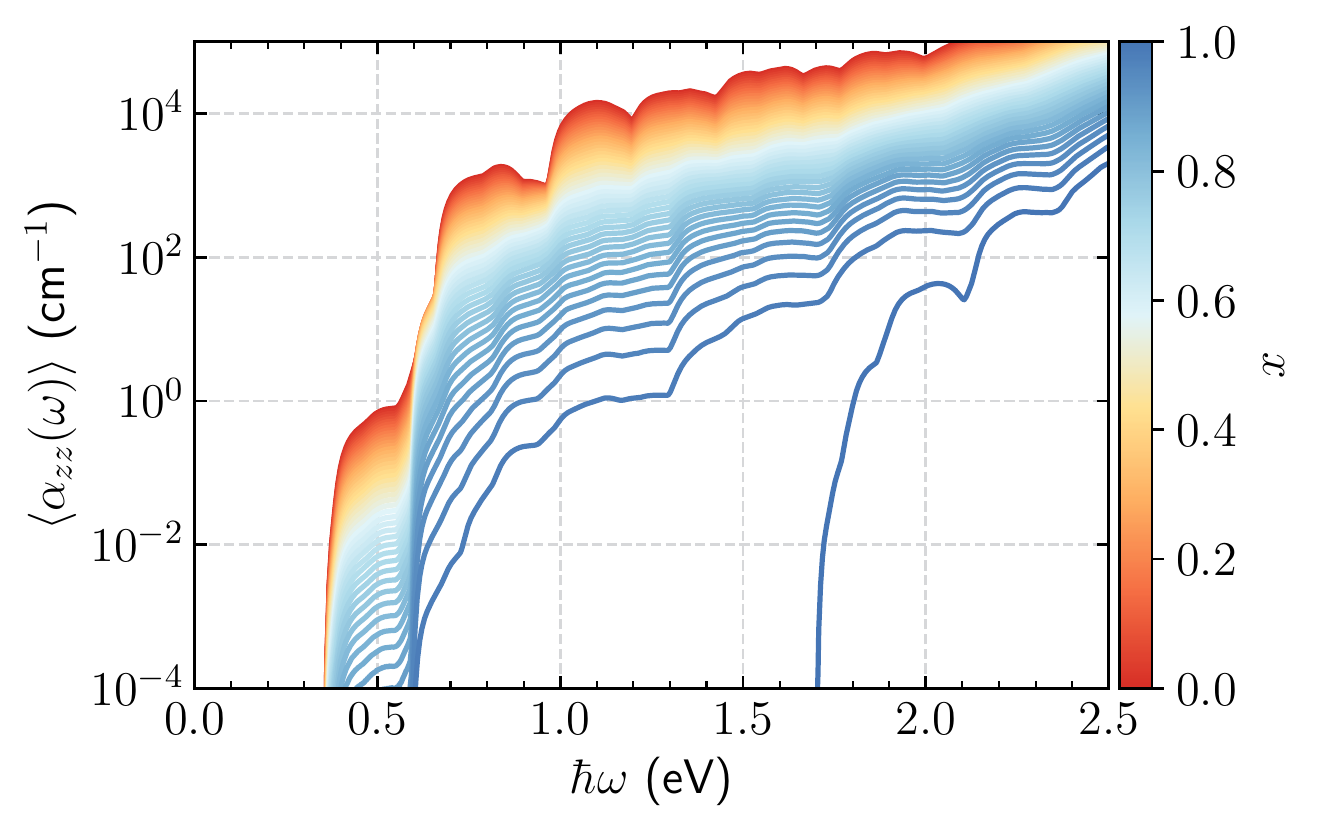} \\
	\end{tabular}   %
   \caption{In-plane (left) and out-of-plane (right) absorption coefficient using 4-atom cells and the bump function as broadening. The curves presented correspond to the absorption coefficient of all individual clusters (top panel), an average of the absorption coefficient over the clusters with the same stoichiometry $n_j/n$ (middle panel), and the SRS-averaged absorption coefficient (bottom panel).}
   \label{fig:4atom-absorption-bump-log}
\end{figure*}

The absorption coefficient at the onset of the spectrum is displayed in Fig.~\ref{fig:4atom-absorption-bump-log}. Here, the results for the 4-atom cells with refined $\kv$-point sampling are shown (top panels). The alloy average obtained from the 8-atom cells for a larger energy range, but using a coarser $\kv$-point grid, is provided the Supp.\ Mat. As there are only 8 inequivalent 4-atom cluster cells, we can compare the absorption coefficients computed for the individual clusters with the SRS average in Fig.~\ref{fig:4atom-absorption-bump-log}. Thanks to representing the peaks in the imaginary part of the dielectric function by the bump \eqref{eq:bump}, which drops to exactly zero and does not feature the long tails of the Lorentzians, the absorption onset is clearly visible as sharp edge in the logarithmic plots of the absorption coefficient.

In the thermodynamic average within the SRS (or GQCA) according to Eq.~\eqref{eq:absorption-coefficient}, all clusters contribute with weights $x_j\neq0$ at any concentration $x$, unless for $x=0$ or $x=1$. For instance, even for $x$ close to one, the pure hex-Ge cluster with its low band gap will have non-vanishing weight. This results in a seemingly unphysical evolution of the absorption edge with alloy composition (see bottom panel of Fig.~\ref{fig:4atom-absorption-bump-log}), where the onset evolves very slowly with increasing Si concentration until it jumps to the absorption edge of hex-Si for $x=1$. This raises the question how to correctly determine the absorption onset. One possibility is to fix a value for a threshold absorption $\alpha_0$, trace for all $x$ at which energy $\hbar\omega$ the absorption rises above this threshold, $\langle\alpha(\omega)\rangle>\alpha_0$, and define the threshold energy as the position of the absorption onset. Another approximate approach is to average only over the clusters with stoichiometry $n_j/n=x$ for a given alloy composition $x$ (see middle panels of Fig.~\ref{fig:4atom-absorption-bump-log}), since these clusters will yield the dominant contributions. This provides a more intuitive visualization of the smooth evolution of the absorption onset with alloy composition.


\subsection{Refractive index}

\begin{figure}
   \includegraphics[width=\linewidth]{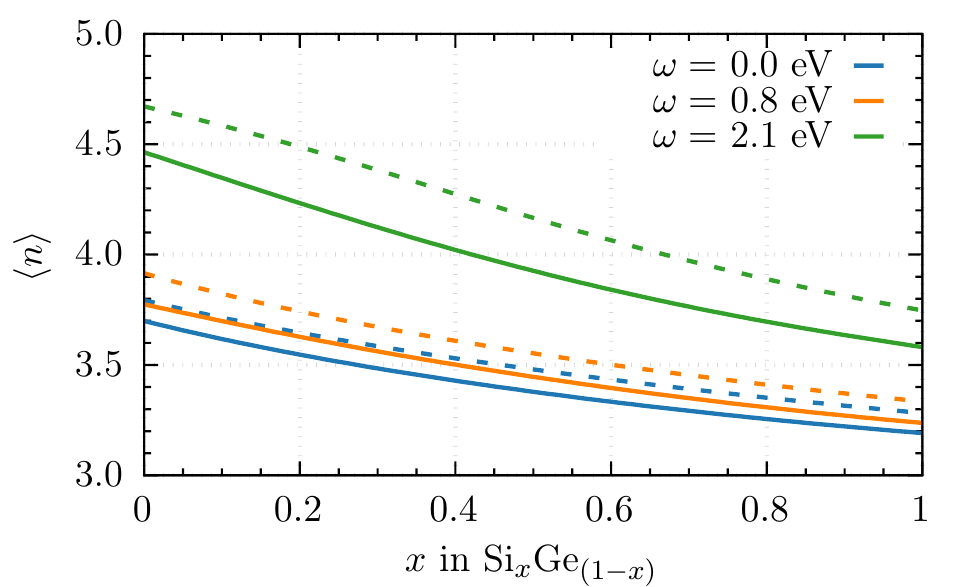}
   \caption{Evolution of the refractive index $\langle n_{ii}(\omega)\rangle$ with alloy composition $x$ at the photon energies $\hbar\omega=0$~\si{\eV}, $0.8$~\si{\eV} (optical fibre communication wavelength), and $2.1$~\si{\eV} (yellow sodium double line). Values for the in-plane (solid lines) and out-of-plane (dashed lines) components are presented, indicating the anisotropy of the material.}
   \label{fig:alloy-refractive-at-0}
\end{figure}

The alloy averages of the frequency-dependent refractive index $n_{ii}(\omega)$, extinction coefficient $\kappa_{ii}(\omega)$, and reflectivity at normal incidence $R_{ii}(\omega)$ are provided for reference in the Supp.\ Mat.

In Fig.~\ref{fig:alloy-refractive-at-0}, we trace the evolution of the refractive index of hex-Si$_x$Ge$_{1-x}$ with alloy composition at fixed frequency. We depict the electronic static refractive index ($\hbar\omega=0)$, along with the refractive index at the technologically important optical fibre communication wavelength which corresponds to photon energies of $0.8$~eV, as well as a photon energy well within the visible spectral range at the wavelength of the yellow sodium double line. In all cases, the refractive index exhibits a monotonous, almost linear evolution with composition $x$. This behavior has also been found experimentally for the cubic alloy~\cite{Schaeffler:2001:InBook}. At $\hbar\omega=0$~\si{\eV}, the refractive index decreases from $3.7$ ($3.8$) for hex-Ge to $3.2$ ($3.3$) for hex-Si for the in-plane (out-of-plane) directions. At $\hbar\omega=2.1$~\si{\eV} (sodium doublet), the refractive index is higher, ranging from $4.5$ ($4.7$) to $3.6$ ($3.8$).


\subsection{Loss function}

\begin{figure}
   \includegraphics[width=\linewidth]{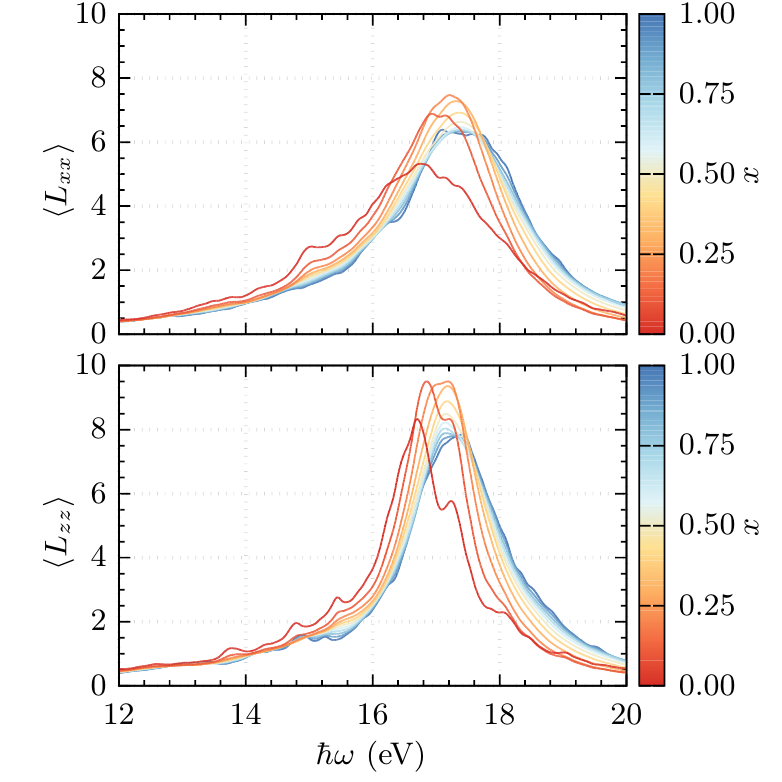}
   \caption{Evolution of the alloy-averaged electron loss function $\langle L_{ii}(\omega)\rangle$ as a function of alloy composition $x$.  Values for both in-plane and out-of-plane components are shown.}
   \label{fig:alloy-eels}
\end{figure}

The energy loss spectra of hex-Si$_x$Ge$_{1-x}$ in the optical limit of vanishing momentum transfer are presented in Fig.~\ref{fig:alloy-eels}. They are obtained from the alloy average of the inverse dielectric functions of the individual clusters according to
\begin{equation}
	\langle L_{ii} (\omega) \rangle = - \Im \langle \varepsilon^{-1}_{ii} (\omega) \rangle. 
	\label{eq:electron-loss}
\end{equation}
We carefully compared this alloy average with the loss function obtained by inverting the alloy-averaged dielectric function and found no significant difference for the resulting spectra (see Fig.~S1). However, we emphasize that we have not taken local-field effects in the evaluation of the loss function into account, as this would result in major additional computational workload for the large number of clusters present in the alloy average.

For low energies, in the spectral range of optical interband transitions, the loss function is dominated by the imaginary part of the dielectric function, $\langle L (\omega) \rangle\approx \Im \langle\varepsilon (\omega)\rangle/\langle\varepsilon(0)\rangle^2$. The plasma frequencies, which correspond to the peak positions in Fig.~\ref{fig:alloy-eels}, can be read from the zeros of the real part of the dielectric function. Due to their very similar density and bonding geometry, the plasma frequencies of the hexagonal compounds, that vary between 16.6~eV for hex-Ge and 17.2~eV for hex-Si, are close to the plasma frequencies of their cubic counterparts (16.0~eV for cub-Ge and 16.9~eV for cub-Si~\cite{monch2001semiconductor}).  The increase of plasma frequency from Ge to Si roughly follows the trend of increasing valence-electron density.

Above the plasma frequency, in the energy range of plasmon oscillations of the $sp$ valence electrons, we observe a pronounced plasmon peak. We have to keep in mind that the positions of the corresponding peak maxima are slightly modified due to the neglect of many-body and local-field effects \cite{Bechstedt:2015:Book}. The tightly bound Ge\,$3d$ semicore electrons do not play an essential role here. This peak shifts towards higher energies from hex-Ge to hex-Si. It also shows significant anisotropy illustrating the different in-plane and out-of-plane bond orientation in the hexagonal crystal structure. 

More refined calculations of loss spectra at finite momentum transfer along with the corresponding experiments are necessary to gain further insight in the local atomic and electronic structure of the alloy.


\section{Transport and thermoelectric properties}
\label{sec:transport}

\subsection{Hexagonal elementary crystals}

\subsubsection{Carrier densities}

\begin{figure}
	\begin{tabular}{c}
	 (a) hex-Ge \\
     \includegraphics[width=\linewidth]{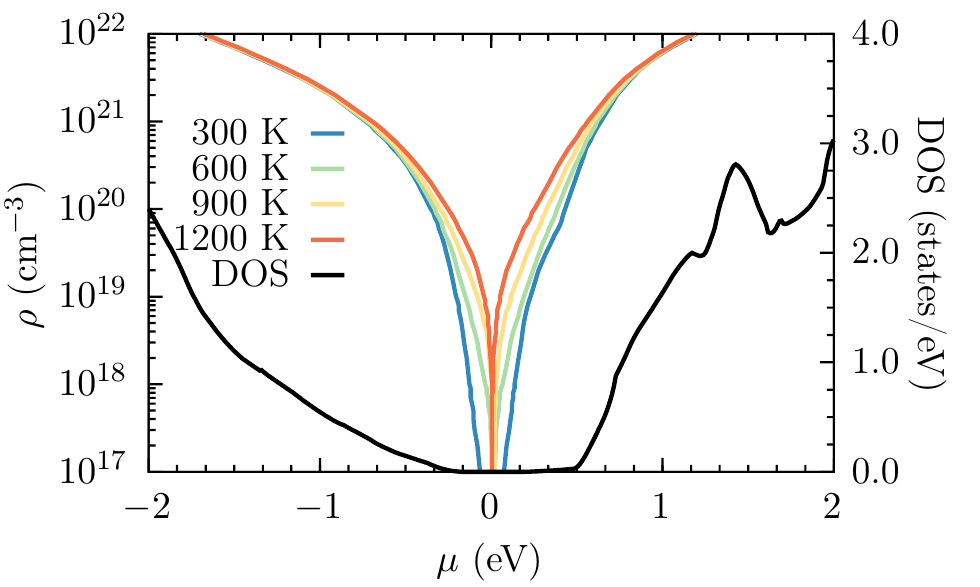} \\
	 (b) hex-Si \\
	 \includegraphics[width=\linewidth]{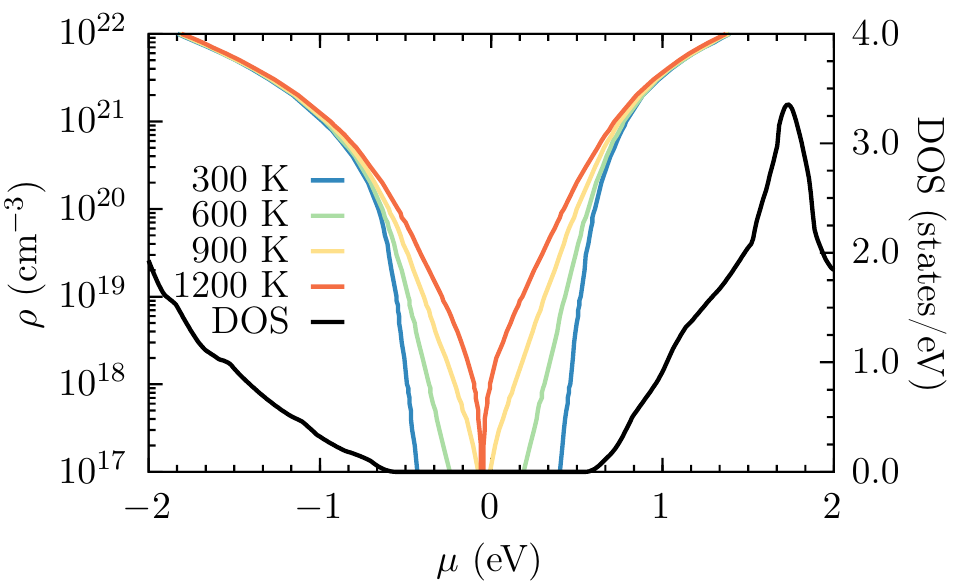} \\
	\end{tabular}
	\caption{Carrier densities as a function of the electron chemical potential $\mu$ and temperature for hexagonal Ge and Si. The respective density of states is superposed as a guide to the eye (black lines). The origin of the scale for the chemical potential is set to the middle of the gap.}
	\label{fig:carrier-density-vs-mu}
\end{figure}

Due to the substrates used to synthesize hexagonal SiGe alloys up to now, the resulting crystals are $n$-doped~\cite{Fadaly:2020:N}. The effect of doping is, of course, relevant for the transport properties and we want to consider it in the following. It is also expected that doping can play a role in stabilizing the hexagonal phase, as suggested by calculations for $n$-doped Ge~\cite{Cai.Deng.ea:2021}.

The carrier density due to doping can be written as
\begin{equation}
\rho =
\begin{cases}
\int\limits^\infty_{\varepsilon_\mathrm{CBM}}d\varepsilon\; f(\varepsilon)\,D(\varepsilon) & \text{for electrons} \\
\int\limits^{\varepsilon_\mathrm{VBM}}_{-\infty}d\varepsilon\; [1-f(\varepsilon)]\,D(\varepsilon) & \text{for holes}
\end{cases},
\label{eq:carrier-density}
\end{equation}
where $\varepsilon_\text{CBM}$ ($\varepsilon_\text{VBM}$) is the conduction-band minimum (valence-band maximum), $f(\varepsilon)=[\exp(\varepsilon-\mu)/(\kB T)+1]^{-1}$ denotes the Fermi distribution for a given chemical potential $\mu$ and temperature $T$, and $D(\varepsilon)$ represents the density of states (DOS).

In Fig.~\ref{fig:carrier-density-vs-mu}, we plot the carrier density of hex-Si and hex-Ge as as a function of the chemical potential for various temperatures. The carrier density varies over several orders of magnitude depending on the position of the chemical potential. When the chemical potential lies within the gap region, the carrier density is determined almost exclusively by the temperature of the electrons, as only the tails of the Fermi distribution reach into energy ranges with non-vanishing DOS and the electron gas can be considered semiclassical.  In $n$- or $p$-doped crystals with shallow donors or acceptors, the chemical potential lies within the bands or in proximity to the band edges. Then, the electron or hole gas becomes degenerate, the carrier density is dominated by the position of the chemical potential, while the influence of temperature is minor, at least for not too low doping.

For the alloys, we further assume that the shallow donor and acceptor levels remain close to the local band edges $\varepsilon_\text{CBM}$ or $\varepsilon_\text{VBM}$. Thus, a more or less homogeneous doping over the entire alloy can be assumed and the condition of thermal equilibrium requires a unique chemical potential. In a more advanced treatment, the chemical potentials of the individual clusters can be used to align their electronic structures and determine the band offsets. 


\subsubsection{Conductivity}

\begin{figure*}
	\begin{tabular}{cc}
			\includegraphics[width=0.33\linewidth]{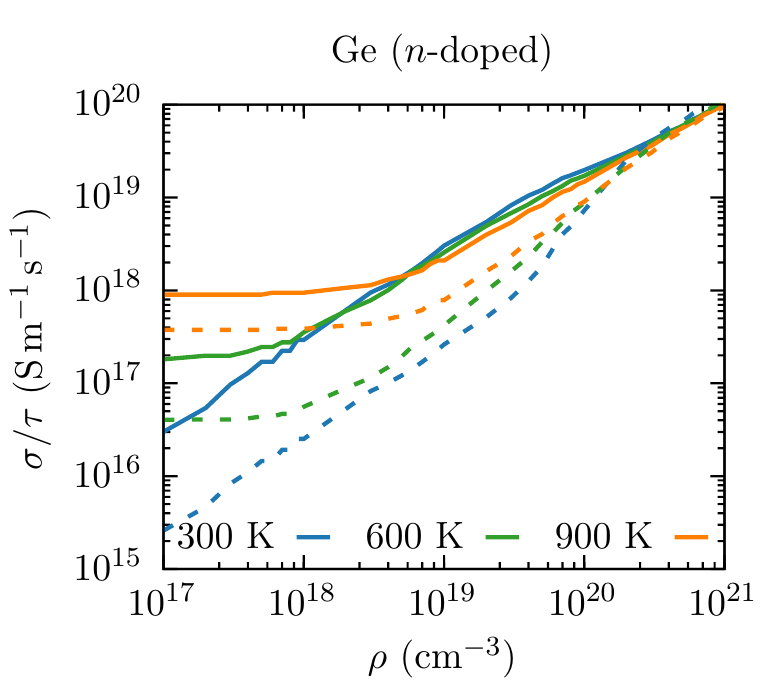} & 
			\includegraphics[width=0.33\linewidth]{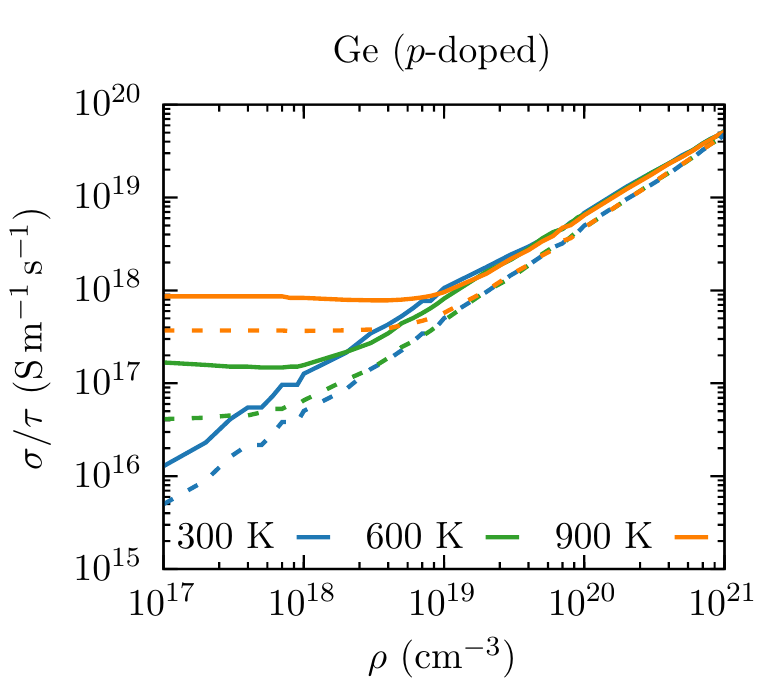} \\
			\includegraphics[width=0.33\linewidth]{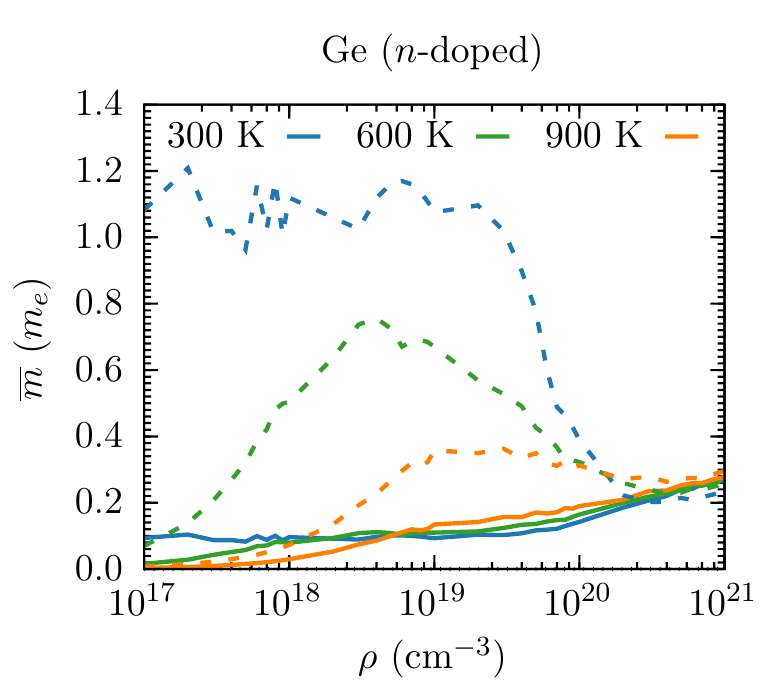} &
			\includegraphics[width=0.33\linewidth]{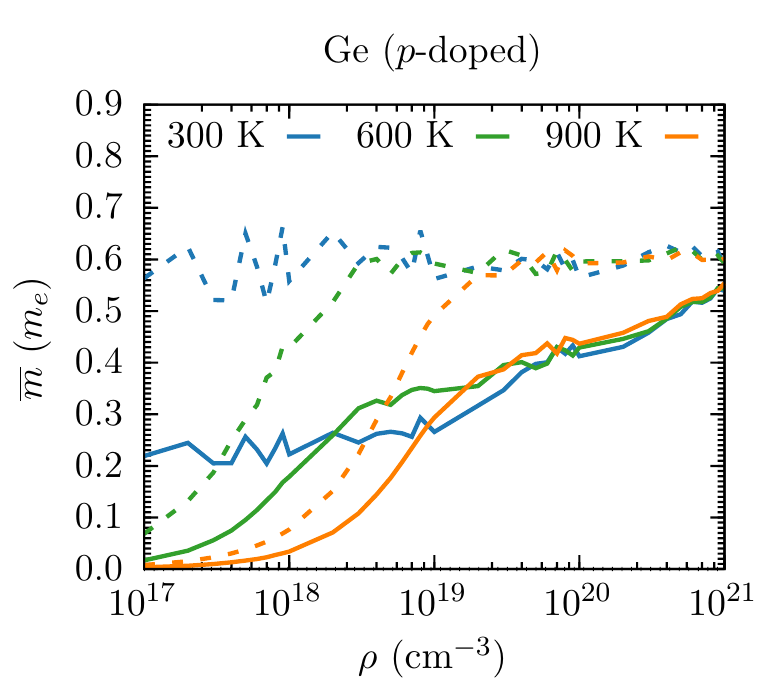} \\
			\includegraphics[width=0.33\linewidth]{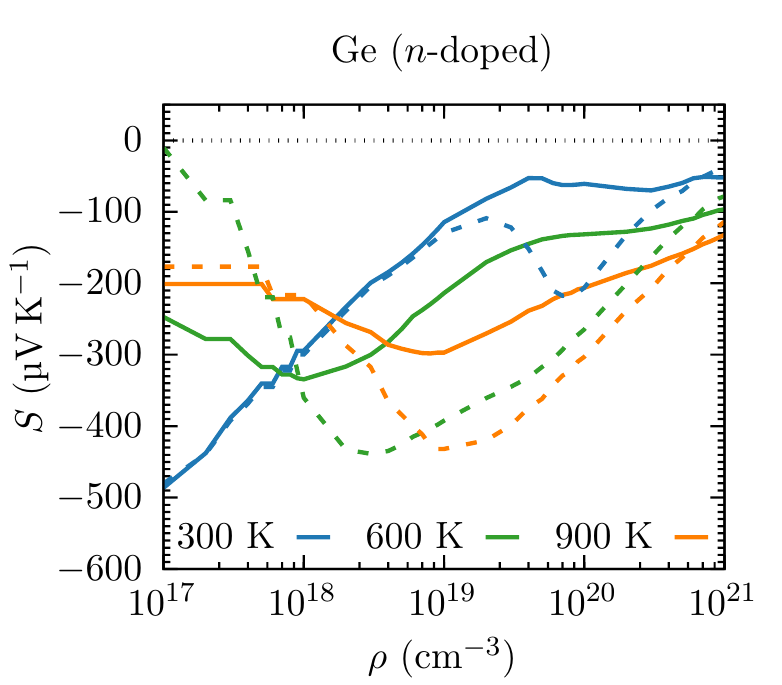} &
			\includegraphics[width=0.33\linewidth]{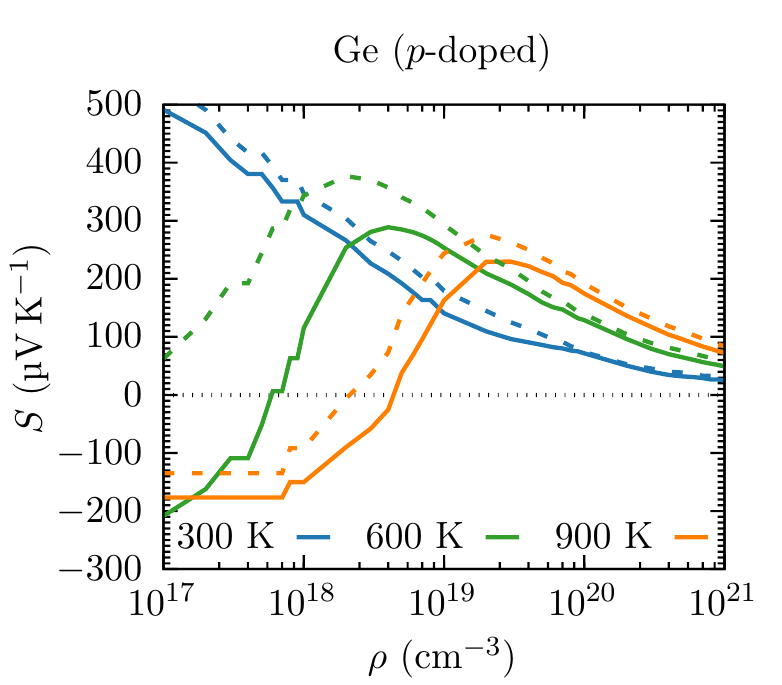} \\
			\includegraphics[width=0.33\linewidth]{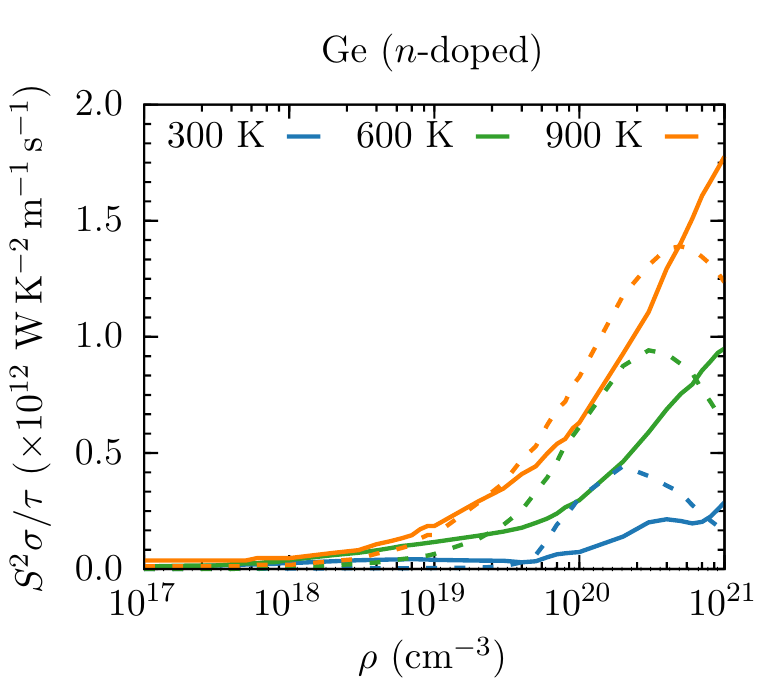} &
            \includegraphics[width=0.33\linewidth]{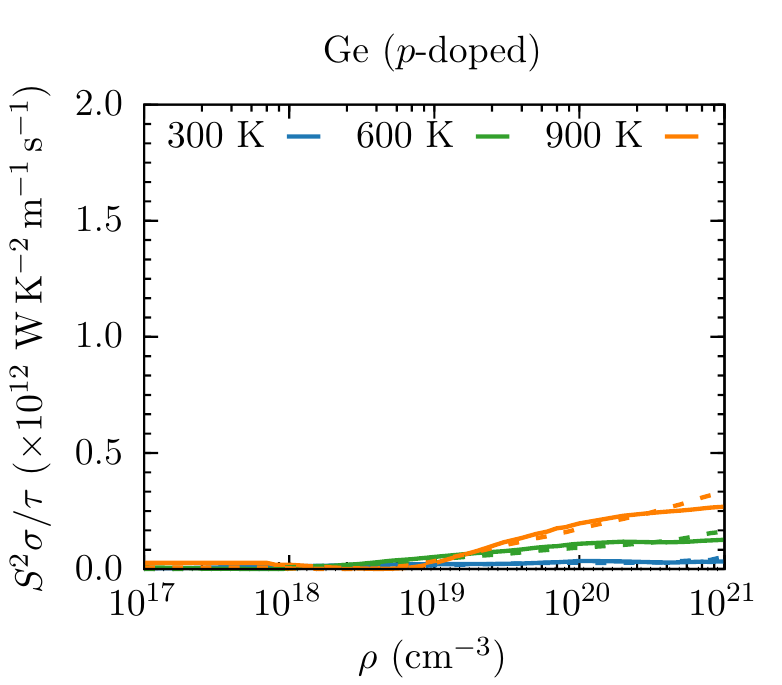} 
	\end{tabular}
	\caption{ Evolution of $\sigma/\tau$, $\overline{m}$, $S$, and $S^2\sigma/\tau$ as a function of temperature and carrier concentration for $n$- and $p$-doped hex-Ge. Solid lines represent the in-plane component of the tensor, while dashed lines indicate the out-of-plane component.}
	\label{fig:bolztrap-ge}
\end{figure*}

\begin{figure*}
	\begin{tabular}{cc}
			\includegraphics[width=0.33\linewidth]{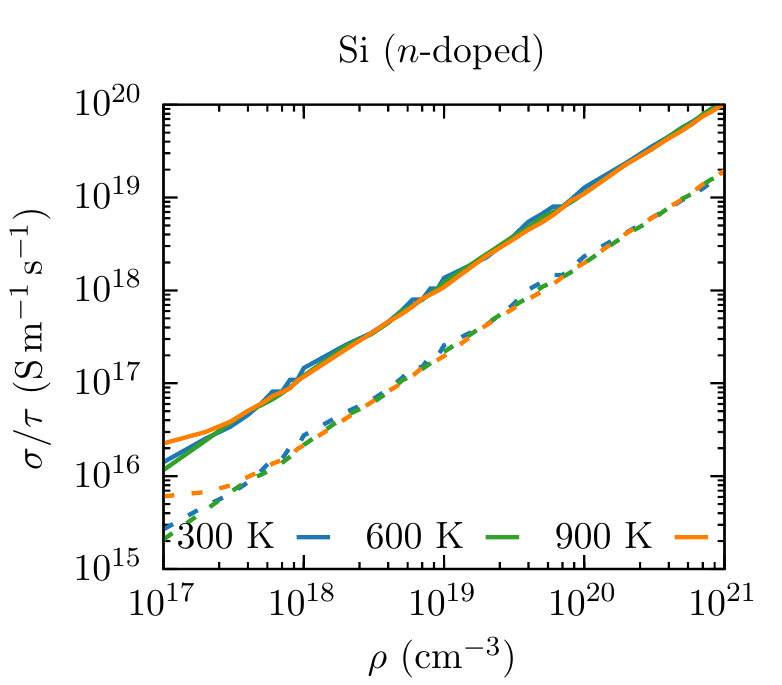} & 
			\includegraphics[width=0.33\linewidth]{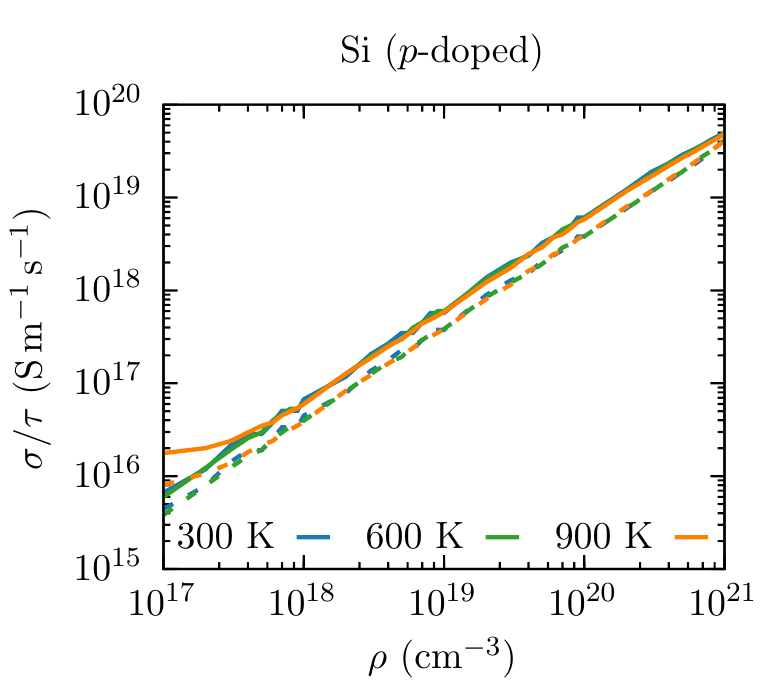} \\
			\includegraphics[width=0.33\linewidth]{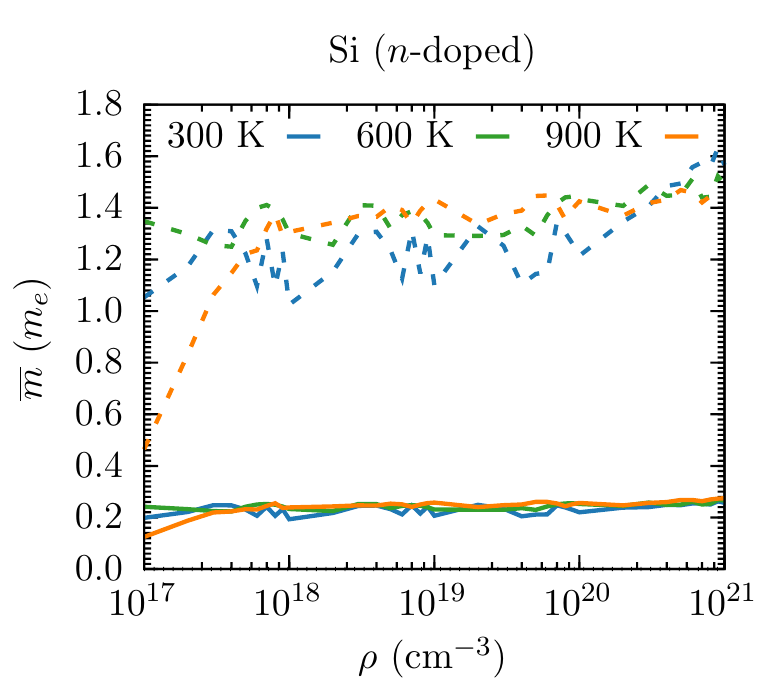} &
			\includegraphics[width=0.33\linewidth]{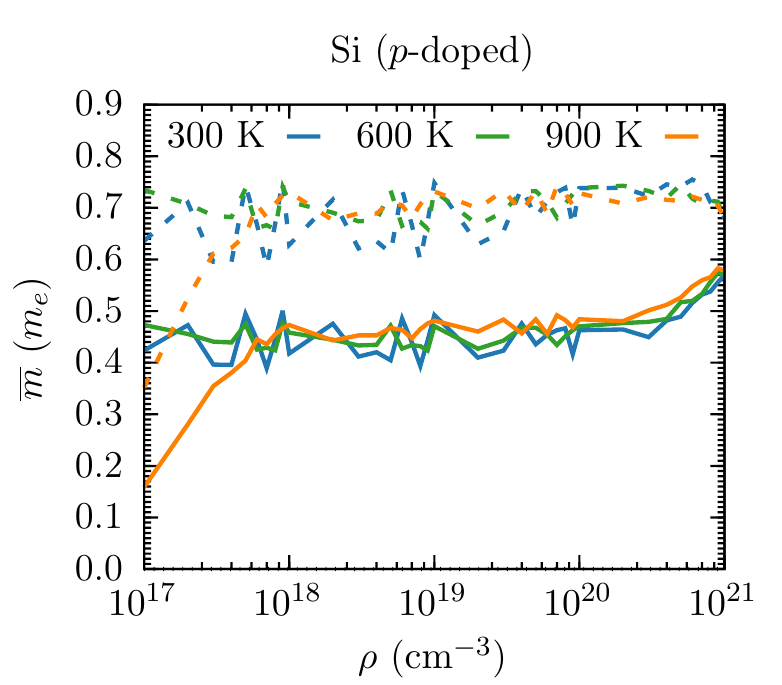} \\
			\includegraphics[width=0.33\linewidth]{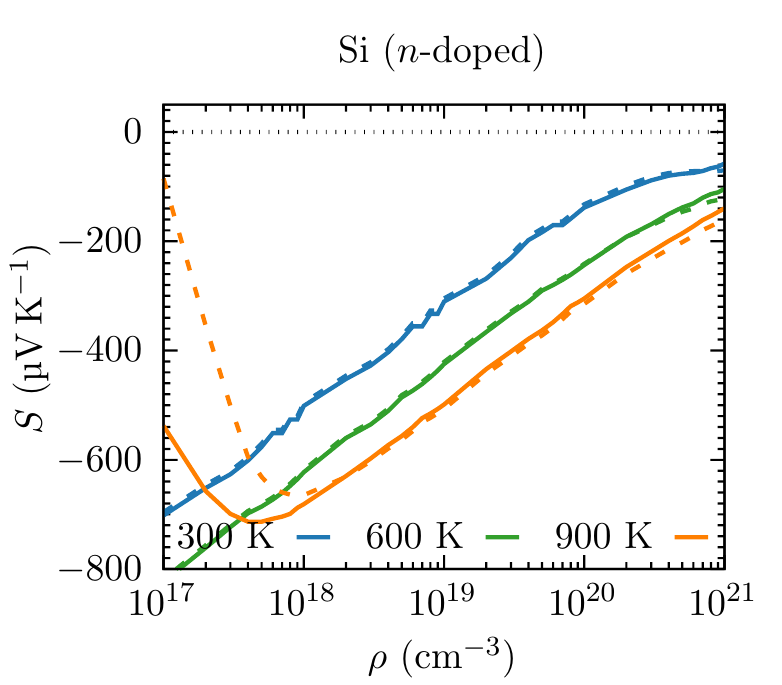} &
			\includegraphics[width=0.33\linewidth]{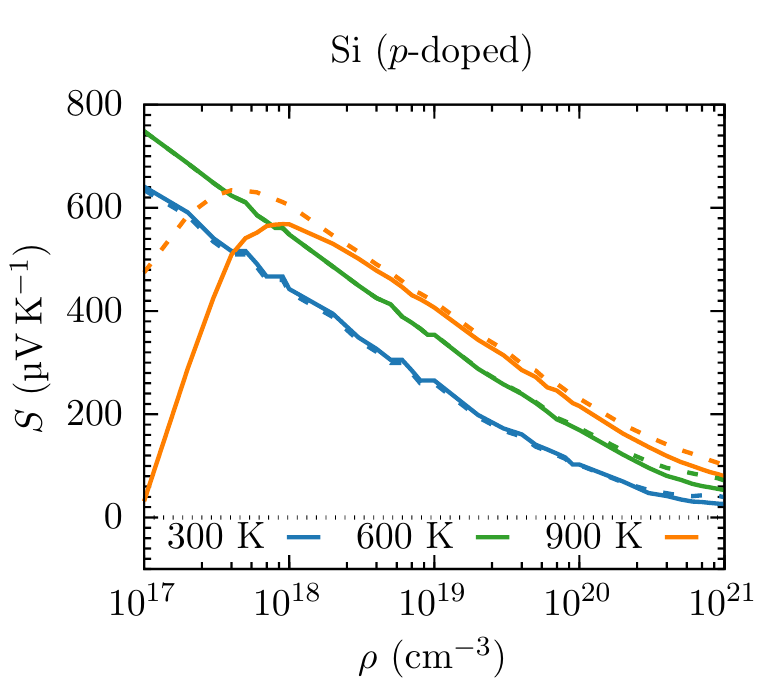} \\
			\includegraphics[width=0.33\linewidth]{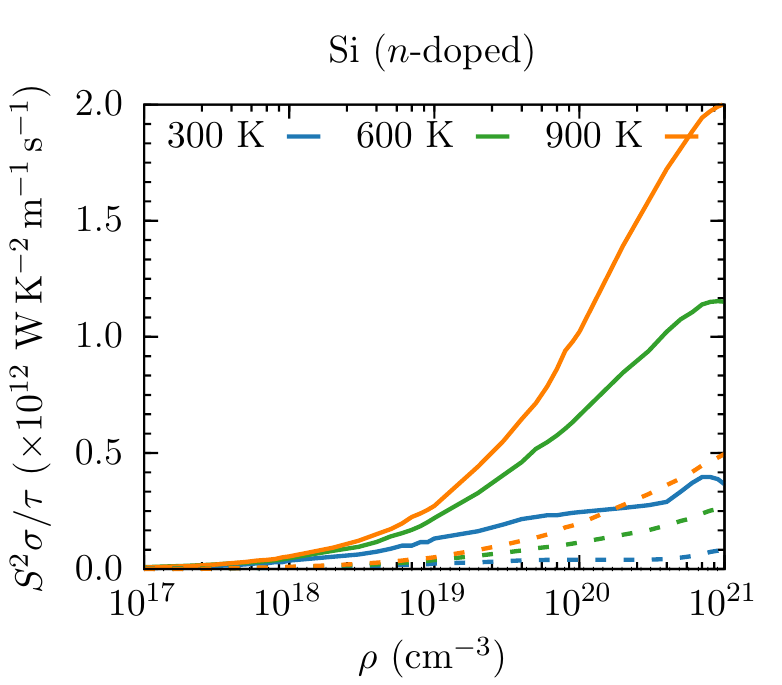} &
            \includegraphics[width=0.33\linewidth]{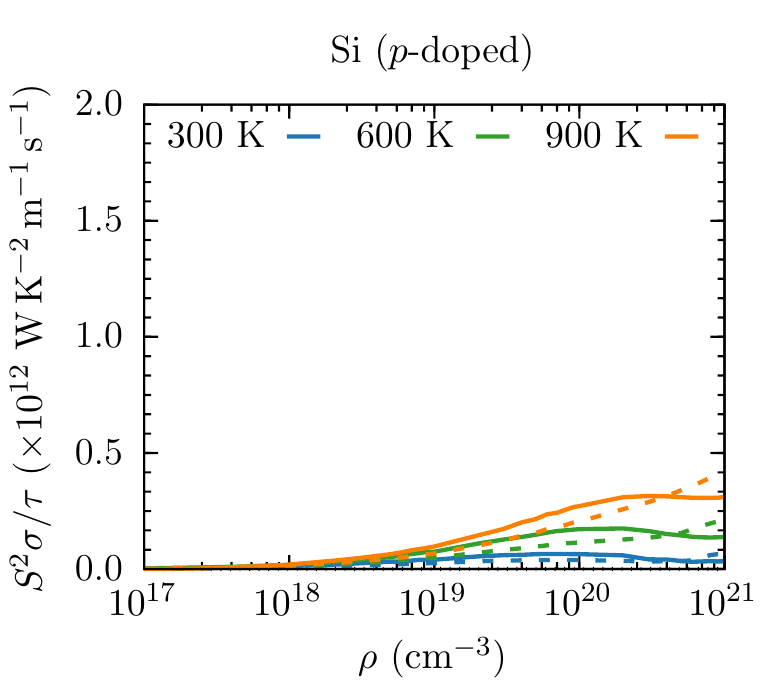} 
	\end{tabular}
	\caption{Evolution of $\sigma/\tau$, $\overline{m}$, $S$, and $S^2\sigma/\tau$ as a function of temperature and carrier concentration for $n$- and $p$-doped hex-Si. Solid lines represent the in-plane component of the tensor, while dashed lines indicate the out-of-plane component.}
	\label{fig:bolztrap-si}
\end{figure*}

Small gradients of the electrostatic potential or the temperature give rise to electrical charge currents and heat currents, respectively. The corresponding response functions are the electrical conductivity $\sigma$ and the thermal conductivity $\kappa$. The non-vanishing components of the electric conductivity tensor  of hex-Ge and hex-Si are presented in Figs.~\ref{fig:bolztrap-ge} and \ref{fig:bolztrap-si} as a function of carrier density and temperature. The electronic contribution to the thermal conductivity $\kappa_\text{e}$ is provided in the Supp.\ Mat. To avoid any bias introduced by the unknown relaxation time, the plots show $\sigma/\tau$ and $\kappa_\text{e}/\tau$.

The dependence of the conductivity on carrier density and temperature is qualitatively very similar. In the case of degenerate carrier gases (i.e., high densities and/or lower temperatures), a temperature-independent linear increase of the conductivity with carrier density is clearly visible, as expected in a Drude picture. In the semiclassical limit (i.e., higher temperatures and/or lower carrier densities), the variations with respect to carrier density are small. However, the conductivity increases with temperature due to the larger number of carriers contributing to the charge transport at higher temperatures. We find similar values of the conductivity for both types of charge carriers. In all cases, the in-plane conductivity is higher than the out-of-plane component.


\subsubsection{Transport masses}

The transport effective masses of both hex-Ge and hex-Si, as plotted in the middle panels of Figs.~\ref{fig:bolztrap-ge} and ~\ref{fig:bolztrap-si}, show a clear anisotropy, with the in-plane masses being considerably smaller than the out-of-plane masses for most of the temperature and carrier-concentration ranges. This fact is in agreement with previous calculations of effective band masses~\cite{Roedl.Furthmueller.ea:2019:PRM}.

In Fig.~\ref{fig:bolztrap-ge}, we can observe that the effective masses of hex-Ge strongly depend on temperature and carrier density. With increasing temperature, a decrease in the effective masses is clearly visible in the low carrier-density region, for both electron and hole masses. Although this behavior is present for all directions, it is more pronounced in the out-of-plane direction. At $T=300$~K and $\rho=10^{19}$~\si{\per\centi\meter\cubed}, the in-plane and out-of-plane electron masses are about $0.1\,m_\mathrm{e}$ and $1.1\,m_\mathrm{e}$, while those for holes are $0.3\,m_\mathrm{e}$ and $0.6\, m_\mathrm{e}$. The in-plane (out-of-plane) masses extracted from band curvatures~\cite{Roedl.Furthmueller.ea:2019:PRM} for the lowest conduction band are $0.09\,m_\mathrm{e}$ and $1.09\,m_\mathrm{e}$ in excellent agreement with the present results for transport masses. For the highest valence band, the situation is slightly more complicated, due to the presence of spin-orbit splitting and heavy and light holes. The masses of the three highest valence bands obtained from the band curvatures are $0.07\,m_\mathrm{e}$, $0.10\,m_\mathrm{e}$ and $0.33\,m_\mathrm{e}$ (in-plane) and $0.53\,m_\mathrm{e}$, $0.12\,m_\mathrm{e}$, $0.05\,m_\mathrm{e}$ (out-of-plane), respectively~\cite{Roedl.Furthmueller.ea:2019:PRM}. These values are consistent with the present findings, although one needs to keep in mind that the transport masses are averaged over the hole-occupied band-structure regions, and do not represent the band curvature at a single point.

In Fig.~\ref{fig:bolztrap-si} we find that the electron masses of Si increase only slightly with the carrier concentration. In addition, they show only a weak dependence on temperature, with the exception of the regime of high temperatures and small carrier concentrations, where a dip in $\overline{m}$ is visible. The values for the electron masses for $T=300$~K and $\rho=10^{19}$~\si{\per\centi\meter\cubed} are around $0.2\,m_\mathrm{e}$ ($1.1\,m_\mathrm{e}$) for the in-plane (out-of-plane) directions, while the hole masses are about $0.5\,m_\mathrm{e}$ ($0.7\,m_\mathrm{e}$).


\subsubsection{Seebeck coefficient}

The Seebeck coefficient $S$, as the proportionality factor between thermoelectric voltage and temperature gradient~\cite{Ashcroft.Mermin:1976:Book}, is shown in Figs.~\ref{fig:bolztrap-ge} and \ref{fig:bolztrap-si}. At room temperature, the absolute value of the Seebeck coefficient varies with carrier concentration, ranging from $500$~\si{\micro\volt\per\kelvin} for hex-Ge and $600$~\si{\micro\volt\per\kelvin} for hex-Si to below  $100$~\si{\micro\volt\per\kelvin}.

Overall, the thermoelectric properties of $n$- and $p$-type hex-Ge and hex-Si are qualitatively similar, due to the similarities in their electronic structure close to the fundamental gap: Both materials have comparably heavy and anisotropic electron masses at the conduction-band minimum, even though these minima are located at different points in the Brillouin zone. Also the three highest valence bands at $\Gamma$ exhibit similar features. The deviating magnitude of the fundamental gaps is of minor importance, since the free-carrier effects are related to intraband transitions. However, there are two qualitative differences: First, since $p$-doped hex-Ge features lower values of $S$, the reduction associated with the temperature increase leads to negative values of $S$ at concentrations below about $10^{18}$~\si{\per\centi\meter\cubed}, depending on temperature). In general, the sign of $S$ is dominated by the sign of the charge carriers. For Ge at high temperature and low hole doping, we find that $S$ is negative, indicating that hole diffusion is not the dominating contribution to the Seebeck effect in this regime, probably due to thermally induced carriers. 

Second, at $300$~\si{\kelvin}, we see a pronounced valley in the out-of-plane Seebeck coefficient for $n$-doped hex-Ge. It is located between  $2\cdot10^{19}$~\si{\per\centi\meter\cubed} and $2\cdot10^{20}$~\si{\per\centi\meter\cubed}, corresponding to approximately the energy range of $0.12$~\si{\eV} to $0.33$~\si{\eV} above the conduction band minimum. The end of this range is close to the second conduction band of hex-Ge, which lies $0.31$~\si{\eV} above the conduction-band minimum~\cite{Roedl.Furthmueller.ea:2019:PRM}. 

At $300$~K and with carrier densities of $10^{19}$~\si{\per\centi\meter\cubed}, $n$-doped hex-Ge and hex-Si display Seebeck coefficients of $-305$~\si{\micro\volt\per\kelvin}  ($-310$~\si{\micro\volt\per\kelvin}) and $-129$~\si{\micro\volt\per\kelvin}  ($-114$~\si{\micro\volt\per\kelvin}) for the in-plane (out-of-plane) directions, respectively. For $p$-doped hex-Ge and hex-Si, these values are $260$~\si{\micro\volt\per\kelvin}  ($265$~\si{\micro\volt\per\kelvin}) and  $179$~\si{\micro\volt\per\kelvin} ($140$~\si{\micro\volt\per\kelvin}). Due to the lack of experimental data for the hexagonal polymorphs, we can only compare with values for cubic Ge and Si. In literature, we find experimental values of the same order of magnitude as we calculated. For example, for cub-Ge, values of about $-330$~\si{\micro\volt\per\kelvin} have been measured under normal conditions ($\rho=10^{14}$~\si{\per\centi\meter\cubed})~\cite{Morozova.Korobeinikov.ea:2020:CEC}. \textit{Ab-initio} calculations of $n$-doped cub-Ge gave values close to $-270$~\si{\micro\volt\per\kelvin} ($\rho=1.1 \cdot 10^{19}$~\si{\per\centi\meter\cubed})~\cite{MurphyArmando:2019:BJAP}. 
Calculations on $n$-doped Si nanosheets yielded values in the range of $-300$ to $-500$~\si{\micro\volt\per\kelvin} (depending on the surface) with a similar range close to $300$ to $500$~\si{\micro\volt\per\kelvin} for $p$-doped Si (both with $\rho=1.1\cdot 10^{19}$~\si{\per\centi\meter\cubed} and $T=300$~\si{\kelvin})~\cite{Nakamura:2016:JJoAP}.


\subsubsection{Thermoelectric power factor}

Finally, combining these quantities, we have computed the thermoelectric power factor $S^2 \sigma$, displayed in Figs.~\ref{fig:bolztrap-ge} and \ref{fig:bolztrap-si}. As a general trend, one observes an increase of $S^2 \sigma/\tau$ with rising carrier density until reaching a maximum close to $n=10^{21}$~\si{\per\centi\meter\cubed}. The increase of $S^2 \sigma / \tau$ is significantly larger at higher temperatures, but these gains are compensated in the power factor by the reduction of $\tau$ which evolves at least as $T^{-3/2}$, e.g.\ in the case of (non-polar) optical-phonon scattering~\cite{Hamaguchi:2017:Book}. We remark that the power factor is the numerator of the thermoelectric figure of merit $zT = S^2 \sigma/(\kappa_\mathrm{e} + \kappa_\mathrm{l})$ \cite{goldsmid2010introduction}. In contrast to the figure of merit, the power factor does not depend on the lattice contribution to the thermal conductivity $\kappa_\mathrm{l}$, that is hard to compute from first principles because it requires the knowledge of phonon frequencies, Gr\"{u}neisen parameters, and dispersion relations (see, e.g.\ Ref.~\cite{Gu.Zhao:2018:JAP}). 

\subsection{Hexagonal SiGe alloys}

\begin{figure*}
   \begin{tabular}{cc}
    \includegraphics[width=0.45\linewidth]{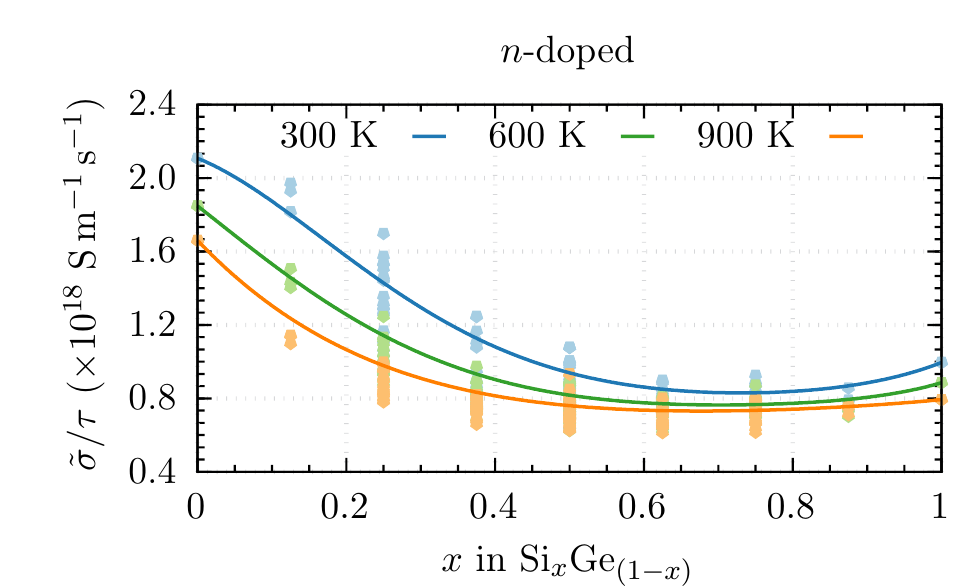} &
    \includegraphics[width=0.45\linewidth]{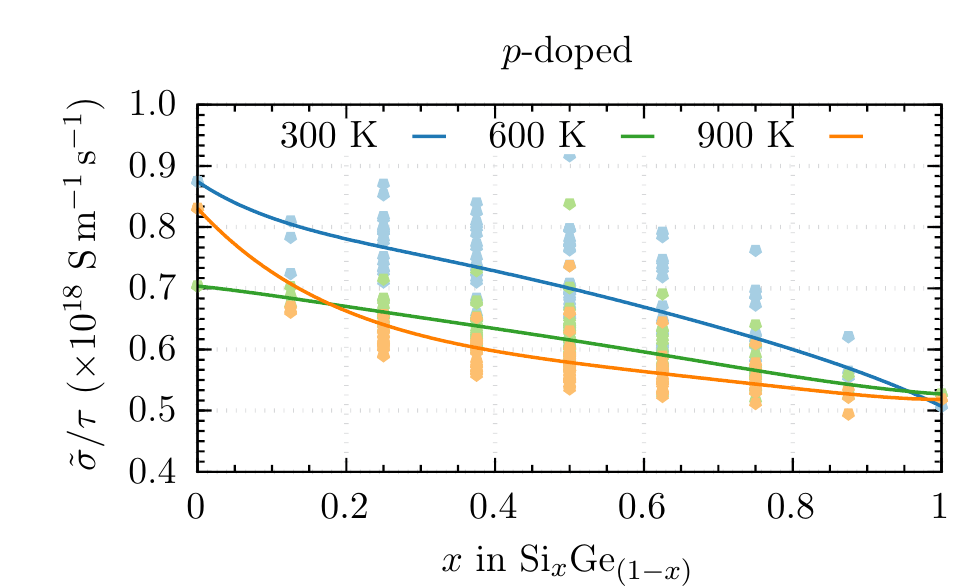}\\
    \includegraphics[width=0.45\linewidth]{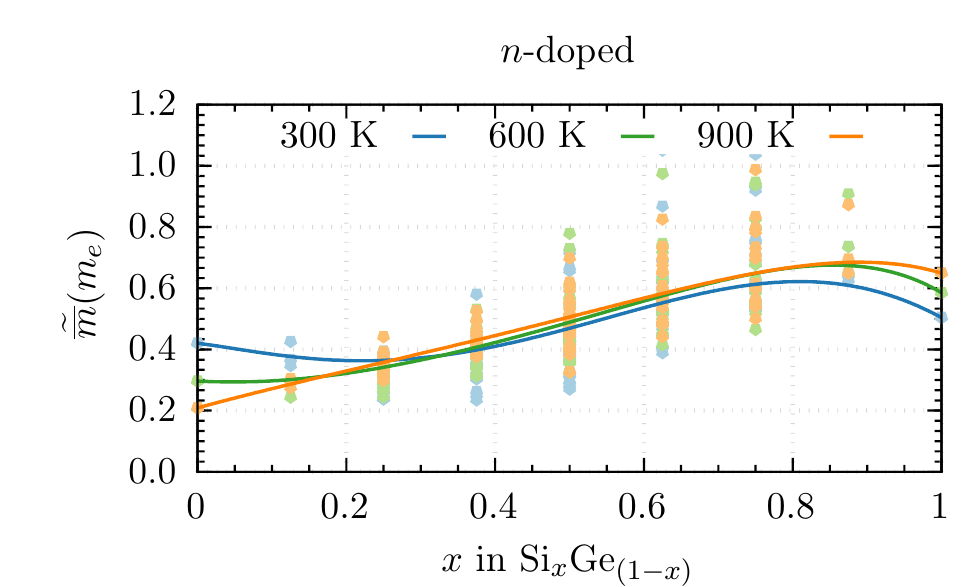} &
    \includegraphics[width=0.45\linewidth]{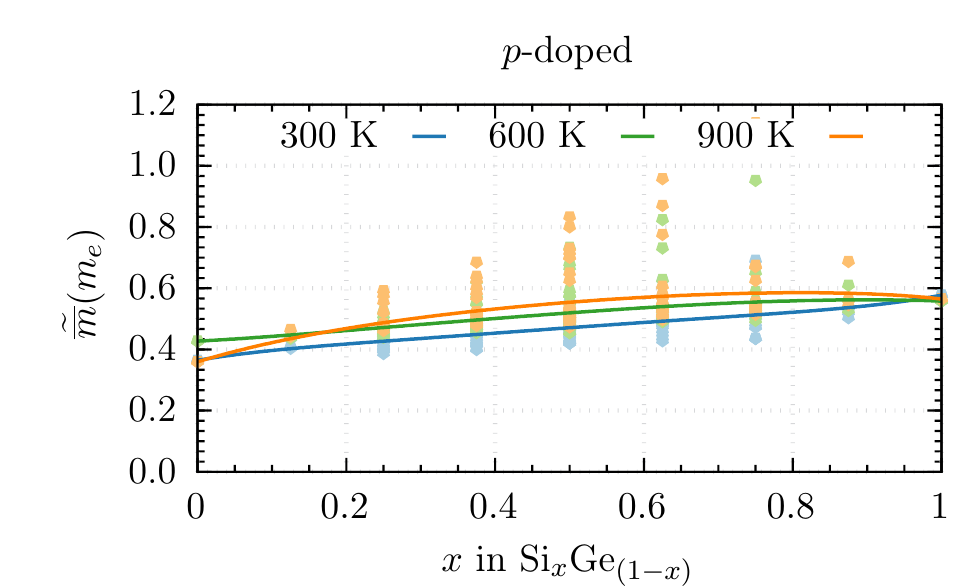} \\
    \includegraphics[width=0.45\linewidth]{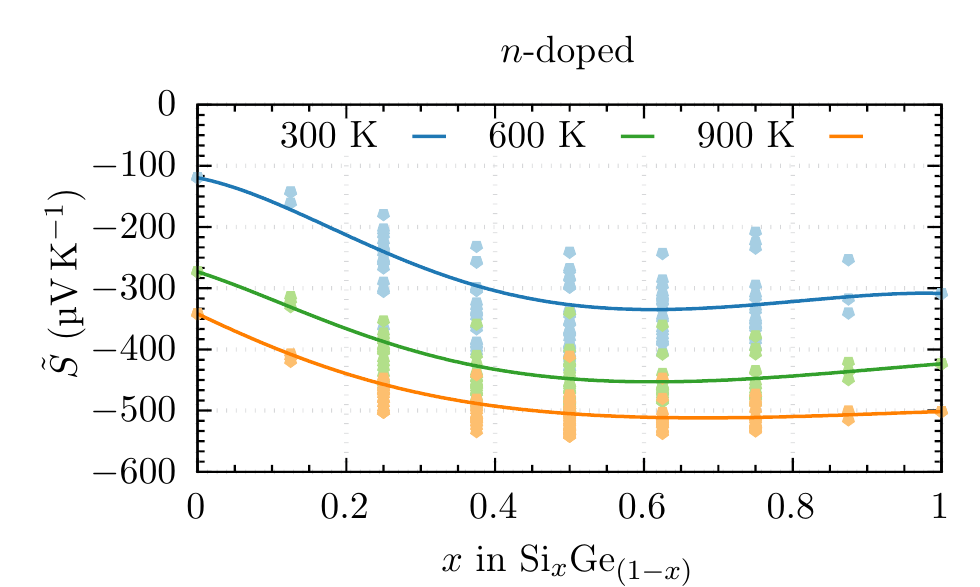} &
    \includegraphics[width=0.45\linewidth]{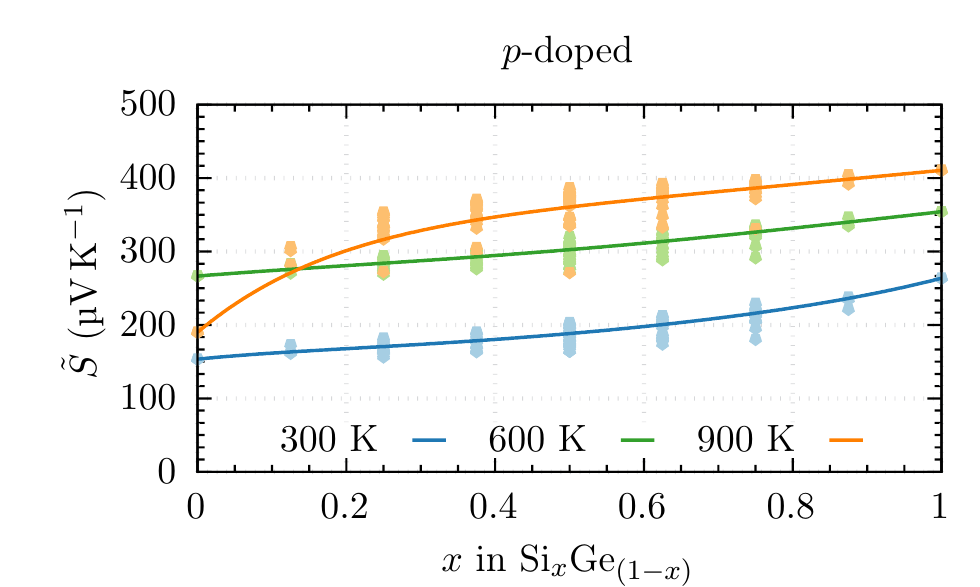} \\
    \includegraphics[width=0.45\linewidth]{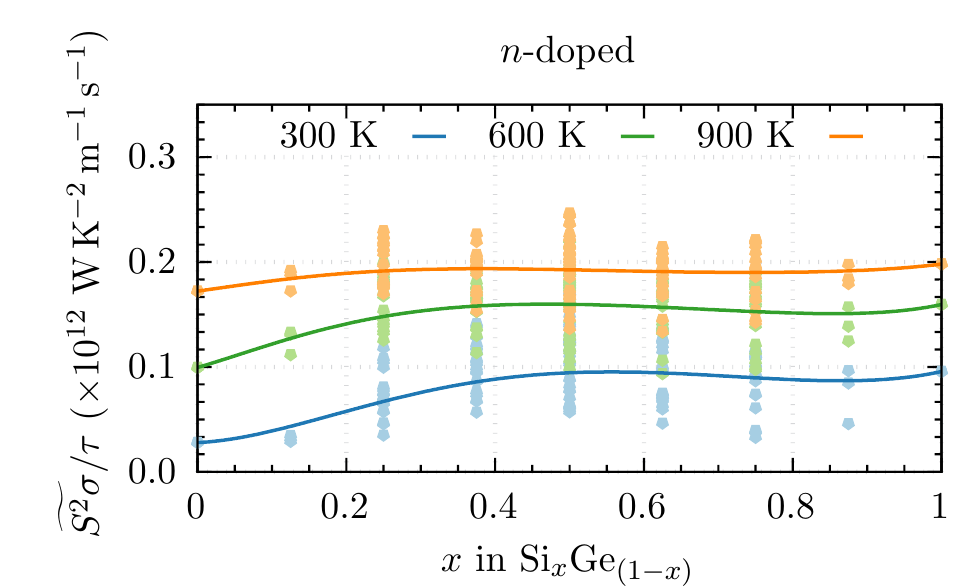} &
    \includegraphics[width=0.45\linewidth]{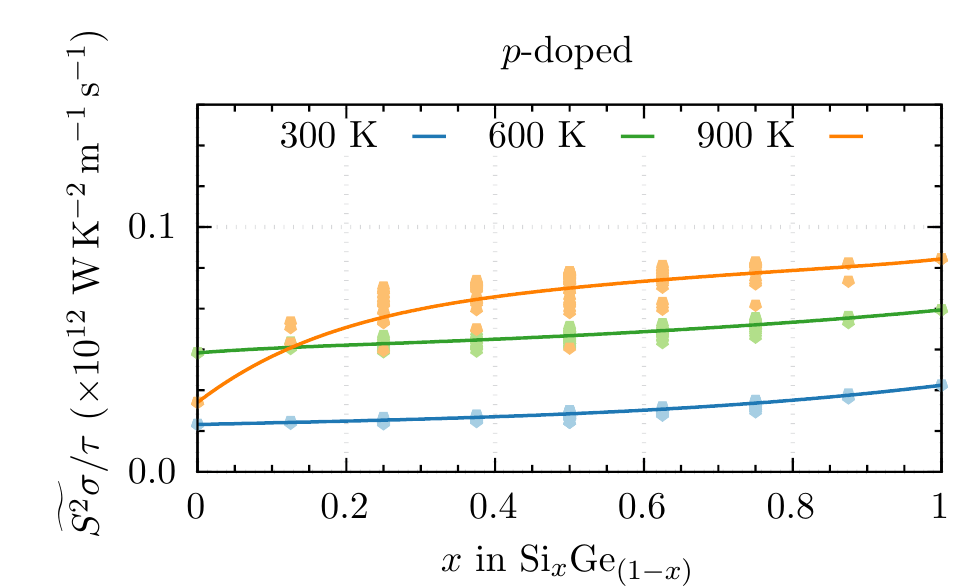}
   \end{tabular}
   \caption{Evolution of the electrical conductivity (over relaxation time), effective transport masses, Seebeck coefficient and power factor (over relaxation time) as a function of Si concentration at different temperatures, for $n$ and $p$ doping, with carrier concentrations of $\rho = 10^{19}$~\si{\per\centi\meter\cubed}. All quantities are alloy-averaged using the SRS model. We plot the traces of the corresponding tensors, thus showing spatially averaged quantities.}
   \label{fig:alloy-transport-doped}
\end{figure*}

We now focus on the evolution of the thermoelectric properties of hex-Si$_x$Ge$_{1-x}$ as a function of composition in the SRS approximation in Fig.~\ref{fig:alloy-transport-doped}. Here, we simplify the analysis considering only averages $\tilde{P}$ of the tensor-valued transport coefficients $P_{ij}$ with respect to crystallographic directions. In addition, we present results only for a carrier concentration of $10^{19}$~\si{\per\centi\meter\cubed}, corresponding to what has been measured in experiment~\cite{Fadaly:2020:N}. 

Figure~\ref{fig:alloy-transport-doped} shows that the electrical conductivity $\sigma/\tau$ decreases with increasing Si concentration. This holds for both $n$-type and $p$-type doping, although the curves are qualitatively different. In the case of hole transport, the conductivity varies almost linearly with composition (at least for not too high temperatures). For electron transport, the conductivity tends to saturate for Si-rich alloys. This behavior has been found in many measurements of the electron Hall mobility of cubic Si$_x$Ge$_{1-x}$ samples~\cite{Busch.Vogt:1960:HPA,Glicksman:1958:PR,Amith:1965:PR} and also in some calculations~\cite{Fischetti.Laux:1996:JAP,MurphyArmando:2019:BJAP}. Due to disorder scattering in the random alloy, a minimum of the electrical conductivity near the middle of the composition range is expected. In fact, the resulting composition dependence of the lifetime $\tau=\tau(x)$ will also influence the composition dependence of the electrical conductivity.

The average electron transport mass varies non-linearly with composition, ranging from $0.2\,m_\mathrm{e}$ for pure hex-Ge and $0.7\,m_\mathrm{e}$ for pure hex-Si, in qualitative agreement with averaged band masses. The nonlinear behavior is due to the various competing conduction-band minima at $\Gamma$ and elsewhere in the Brillouin zone that are very close in energy for some SiGe clusters.
The average hole transport mass varies almost linearly between $0.4\,m_\mathrm{e}$ and $0.6\,m_\mathrm{e}$, as the valence-band maxima of all cluster lie at the $\Gamma$ point and are qualitatively identical in terms of their degeneracies. The small bowing of the curve becomes stronger with increasing temperature.

The alloy-averaged Seebeck coefficient for $n$-doping shows a clear bowing, and tends to saturate for high Si concentrations (such as observed for $\sigma/\tau$). In addition, increasing the temperature leads to an overall reduction of the Seebeck coefficient, as expected, reaching values below $-500$~\si{\micro\volt\per\kelvin} at $900$~\si{\kelvin} (for $x>0.5$). For $p$-doping, weaker and almost linear dispersion is observed. Again, higher temperatures lead to higher Seebeck coefficients, reaching $400$~\si{\micro\volt\per\kelvin} at $900$~\si{\kelvin} (but only for very high concentrations of Si). We observe that $p$-doping leads to smaller absolute values of the Seebeck coefficient than $n$-doping.


\section{Summary and Conclusions}
\label{sec:summary}

In this work, we performed first-principles calculations of optical, transport and thermoelectric properties of hexagonal Si$_x$Ge$_{1-x}$ alloys. The disordered alloy is described by an accurate thermodynamic averaging procedure relying on a cluster decomposition which takes all possible 8-atom SiGe configurations into consideration. The structural and electronic properties of the individual atomic arrangements are described by means of DFT with state-of-the-art exchange-correlation functionals. We first demonstrate the miscibility of hexagonal Si and Ge over the entire composition range at room temperature and above. Building up on the electronic structures, composition-dependent frequency-dependent dielectric functions and derived optical quantities are calculated. Moreover, we studied the effect of alloying on transport and thermoelectric quantities by solving the linearized Boltzmann equation within the constant relaxation time approximation. Thus, we provide crucial data for the characterization of this novel materials system regarding potential applications in optoelectronics and photonics. In particular, we investigated the tunability of the absorption edge and the refractive index with alloy composition, thus providing important reference data for future experiments. We hope to spark further experimental and theoretical research which will widen and deepen the knowledge about hexagonal group-IV materials and allow to overcome the limitations inherent to some of the theoretical approaches employed in this work.


\begin{acknowledgments}
We acknowledge funding for the projects SiLAS (Grant Agreement No.\ 735008) and  OptoSilicon (Grant Agreement No.\ 964191) from the European Union’s Horizon 2020 research and innovation program.
P.B. acknowledges financial support from the CFisUC through the project UIDB/04564/2020 and FCT under the contract 2020.04225.CEECIND. Computing time was granted by the Leibniz Centre on SuperMUC (Grant No.~pr62ja).
\end{acknowledgments}


%

\end{document}